\documentclass[letterpaper]{article} 
\usepackage[draft]{aaai2026}  

\usepackage[T1]{fontenc}
\usepackage{textcomp}
\usepackage{lmodern}
\usepackage{helvet}
\usepackage{courier}

\usepackage[hyphens]{url}  
\usepackage{graphicx} 
\usepackage{amsmath,amssymb}
\usepackage{enumitem}
\usepackage{booktabs}
\usepackage{subcaption}

\usepackage{adjustbox}

\usepackage{natbib}  
\usepackage{caption} 
\usepackage{amsmath, amssymb} 
\usepackage{algorithm}
\usepackage{algorithmic}
\newcommand{\whisper}{Whisper}
\newcommand{\qwen}{Qwen2-Audio}
\usepackage{newfloat}
\usepackage{listings}
\usepackage{bibentry}
\DeclareCaptionStyle{ruled}{labelfont=normalfont,labelsep=colon,strut=off} 
\floatstyle{ruled}
\newfloat{listing}{tb}{lst}{}
\floatname{listing}{Listing}
\usepackage{xcolor}
\definecolor{turquoise}{RGB}{64,184,188}

\title{Beyond Transcription: Mechanistic Interpretability in ASR}
\author{
  Neta Glazer,
  Yael Segal-Feldman,
  Hilit Segev,
  Aviv Shamsian,
  Asaf Buchnick,
  Gill Hetz,
  Ethan Fetaya,
  Joseph Keshet,
  Aviv Navon
}
\affiliations{
   aiOla Research\\
  neta.glazer@aiola.com
}

\begin{document}
\maketitle

\begin{abstract}
Interpretability methods have recently gained significant attention, particularly in the context of large language models, enabling insights into linguistic representations, error detection, and model behaviors such as hallucinations and repetitions. However, these techniques remain underexplored in automatic speech recognition (ASR), despite their potential to advance both the performance and interpretability of ASR systems. In this work, we adapt and systematically apply established interpretability methods such as logit lens, linear probing, and activation patching, to examine how acoustic and semantic information evolves across layers in ASR systems. Our experiments reveal previously unknown internal dynamics, including specific encoder-decoder interactions responsible for repetition hallucinations and semantic biases encoded deep within acoustic representations. These insights demonstrate the benefits of extending and applying interpretability techniques to speech recognition, opening promising directions for future research on improving model transparency and robustness.

\end{abstract}


\section{Introduction}

Automatic Speech Recognition has advanced significantly in recent years, largely driven by powerful neural architectures trained on extensive speech datasets~\cite{radford2023robust,Qwen-Audio}. Modern ASR systems commonly utilize encoder-decoder transformer architectures, facilitating robust recognition across diverse languages, accents, and acoustic conditions~\cite{radford2023robust,chu2024qwen2,abouelenin2025phi}. Recently, architectural approaches have diverged, with some models adopting Large Audio-Language Models (LALMs) that integrate pretrained language models~\cite{tang2023salmonn,das2024speechverse,abouelenin2025phi}, such as Qwen2-Audio \cite{Qwen-Audio,chu2024qwen2}, while others continue to rely on transformers specifically trained for speech, such as Whisper~\cite{radford2023robust}.


In parallel, interpretability has become a central research focus in large language model (LLMs)~\cite{brown2020language,touvron2023llama,team2024qwen2}, with growing efforts to understand how these models represent information and produce decisions~\cite{Luo2024FromUT}.  Techniques linear probing \cite{belinkov2022probing,mckenzie2025detecting}, logit lens \cite{geva2022transformer,nostalgebraist2020logitlens}, and activation patching \cite{meng2022locating,wang2022interpretability} allow researchers to reverse-engineer internal model behavior, revealing the structure of linguistic representations and tracing the origins of specific outputs. These tools have proven crucial in diagnosing failure modes like hallucinations and reasoning errors, and have contributed to improving model safety and reliability.

In this work, we take a first step toward bridging the interpertability gap in ASR. We examine the internal behavior and dynamics of modern ASR and Large Audio-Language Models to understand the mechanisms behind key error phenomena such as hallucinations, repetition loops, and contextually biased outputs. In addition, we trace how predictions evolve across layers, identify which components drive specific decoding behaviors, and reveal how contextual expectations compete with acoustic evidence. Beyond error analysis, we investigate the rich representations these models encode, from quality prediction signals embedded in decoder states to localized attention mechanisms that control model failures.

To better understand these phenomena, we systematically adapt interpretability techniques to reveal the internal mechanisms of these models. We find that acoustic and semantic attributes are linearly decoded in the encoder layers, with clearer separation in upper layers. We discover that hallucination-related signals are strongly expressed in the decoder's residual stream, enabling an accurate real-time quality prediction. Furthermore, we identify the specific mechanisms responsible for repetitions, revealing which components control these failure modes. Additionally, we show that contextual bias emerges within the encoder itself and can override acoustic evidence, challenging assumptions about encoder-decoder role separation.

This work takes a step toward systematically understanding of the internal dynamics and decision-making processes of speech recognition models, opening new directions for improving their reliability and performance.

\section{Related Work}







A substantial amount of research has focused on understanding how LLMs process and represent information \cite{rauker2023toward}, ranging from identifying specific circuits responsible for particular tasks \cite{hanna2023does, goldowsky2023localizing} to exploring how the model ``thinks'' \cite{schut2025multilingual} and which components responsible for repetitions \cite{yona2025interpreting,barbero2024transformers}. The methods used to understand the model are varied. For example, the logit lens and its improved variant Tuned Lens track how token predictions evolve across layers, providing a layer-wise view of model behavior \cite{nostalgebraist2020logitlens, geva2022transformer, belrose2023eliciting}. Complementary approaches such as linear probing test whether models encode features like syntax or factual knowledge in directions recoverable by simple classifiers \cite{belinkov2022probing, mckenzie2025detecting, hernandez2023linearity}. Building on this, activation patching and causal tracing explore the causal role of specific hidden states by swapping or ablating them to observe changes in outputs \cite{meng2022locating, heimersheim2024use}. More recently, attribution patching has extended these ideas to finer-grained structures by using gradients to pinpoint influential neurons \cite{syed2023attribution, nanda2023attribution, kramar2024atp}. Collectively, these methods represent diverse attempts to interpret how LLMs operate internally.

Several studies have examined the internal representations learned by the \whisper{} model.
These studies show that the Whisper encoder captures noise‑related features, speaker identity, and emotional content \cite{gong2023whisper, upadhyay2024layer, zhao2024whisper}, while the decoder also encodes speaker traits and reacts to language shifts \cite{berns2023speaker}. However, these works did not target model interpretability. A blog post by \citet{ellena2023interpreting} offers the first large-scale interpretability analysis, revealing that encoder neurons align with human-interpretable phoneme patterns, and that the decoder mainly acts as a weak language model. Other works extend this line of inquiry in different directions: \citet{lioubashevski2024looking} 
show that decoder based LLM including \whisper{} first stabilize on the top-ranking token, then successively the next highest-ranked tokens; \citet{ballier2024probing} apply probing methods to analyze calibration curves across multiple languages; and \citet{baranski2025investigation} investigate hallucinations over non-speech segments, aiming to catalog frequently occurring hallucinations rather than localizing them within model components. \citet{yang2024prompts} explores the influence of text prompts on Whisper's outputs.

\section{Method}
\label{sec:methods}
In this section, we present the interpretability techniques employed in our study. Since these methods were originally developed for LLMs or vision models, we describe the adaptations required to apply them effectively in the ASR setting. We begin by introducing the notation used throughout.

\subsection{Preliminaries and Notation}
\label{sec:notation}
We consider encoder–decoder ASR models that generate a sequence of tokens \( \mathbf{y} = (y_1, \dots, y_T) \) from input audio \( \mathbf{x} \), using a Transformer encoder and decoder~\cite{vaswani2017attention}. Let \( L_e \) and \( L_d \) denote the number of encoder and decoder layers, respectively, and \( d \) the hidden dimension.

The encoder processes audio into a sequence of hidden vectors. We denote by \( \mathbf{h}^{l_e} \in \mathbb{R}^{F \times d} \) the encoder representation at layer \( l_e \in \{1, \dots, L_e\} \), where \( F \) is the number of audio frame representations after feature extraction. We use \( \mathbf{h}_\tau^{l_e} \in \mathbb{R}^d \) to refer to the representation at position \( \tau \in \{1, \dots, F\} \) and by \( \mathbf{h}_t^{l_d} \in \mathbb{R}^d \) the decoder hidden state at layer \( l_d \) and token position \( t \). The decoder output is projected to vocabulary logits using the unembedding matrix  \( E \in \mathbb{R}^{|\mathcal{V}| \times d} \) using
\[
\mathbf{z}_t^{l_d} = E \cdot \mathbf{r}_t^{l_d} \in \mathbb{R}^{|\mathcal{V}|}.
\]
Here \( \mathbf{r}_t^{l_d} \in \mathbb{R}^d \) is the residual stream, which captures the decoder's intermediate representation after layer normalization but before output projection.

\subsection{Interpretability Methods}

\paragraph{Logit Lens.}

The logit lens \cite{geva2022transformer} provides a layer-by-layer view of how the model's predictions evolve during decoding. At each decoding step \( t \), we take the residual stream \( \mathbf{r}_t^{l_d} \) from each decoder layer \( l_d \), and project it to the vocabulary space using the unembedding matrix \( E \), to produce the logits vector $\mathbf{z}_t^{l_d}$. 
We extract the top-\(k\) tokens from each \( \mathbf{z}_t^{l_d} \) to analyze how predictions develop across layers.
To quantify this process, we follow \citet{geva2022transformer,lioubashevski2024looking}, and define the \textit{saturation layer} of a token \( t \) as the earliest decoder layer whose top-1 prediction matches the final output and remains stable:
\[
l^*_t = \min \left\{ l_d : \arg\max \mathbf{z}_t^{l_d} = \arg\max \mathbf{z}_t^{L_d} \right\}.
\]
This provides insight into when the model effectively commits to a prediction.

\paragraph{Activation Probing.}
\label{sec:probing}
Probing tests whether specific attributes are encoded in a model's hidden representations~\cite{belinkov2022probing}. We use linear probes: simple classifiers trained on frozen activations \( \mathbf{h} \in \mathbb{R}^d \) to predict a label, 
\[
\mathcal{P}(\mathbf{h}) = W\mathbf{h} + b,
\]
where \( W \in \mathbb{R}^{k \times d} \), \( b \in \mathbb{R}^k \) are trained using cross-entropy or regression loss. High accuracy suggests that the attribute is linearly decodable from the representation~\cite{hernandez2023linearity}.

For decoder, we probe token-level hidden states (typically at the final position). In the encoder, where representations are aligned with audio frames, we average across time to produce a fixed-length vector. Probes may be reused at inference to monitor internal structure with minimal overhead~\cite{mckenzie2025detecting}.

\paragraph{Intervention-Based Analysis.}
\label{sec:patching_methods}
Causal intervention methods study why a model produces a particular output by modifying its internal activations and observing the effect on predictions. If modifying a component changes the output, that component is said to play a causal role in the behavior. This idea underlies recent work on factual editing and mechanism tracing in LLMs and vision models~\cite{meng2022locating, wang2022interpretability, ben2024lvlm, haklay2025position}.  
We adapt two standard interventions, component patching and ablation, to analyze \whisper{} and \qwen{}.

\emph{Component patching.} In this technique, we run the model on two inputs: a target input and a reference input. During the forward pass on the target input, we replace the activation of a selected component with the one recorded from the reference input. 
Formally, let \( \mathbf{a}_C^{\text{orig}} \) be the activation at component \( C \) when running the original input \( \mathbf{x}_{\text{orig}} \), and let \( \mathbf{a}_C^{\text{ref}} \) be the corresponding activation from a reference input \( \mathbf{x}_{\text{ref}} \). We compute a patched activation as:
\[
\tilde{\mathbf{a}}_C = (1 - \alpha)\, \mathbf{a}_C^{\text{orig}} + \alpha\, \mathbf{a}_C^{\text{ref}}, \quad \alpha \in \mathbb{R}_+.
\]

In our experiments, we use white noise as the reference input, which serves to disrupt the natural computation. This helps reveal components that are critical for maintaining acoustic fidelity or 
contextual bias.

\emph{Ablation.}  
Ablation tests whether a component is necessary for a model behavior by removing its contribution during inference~\cite{vig2020investigating}.
This is done by zeroing out the activation at component \( C \), $\tilde{\mathbf{a}}_C = \mathbf{0}$,
and observing the change in output.  
If the prediction is degraded or altered, we interpret this as evidence that \( C \) is important for producing the original behavior.

\emph{Intervention Scope.}  
We apply interventions on both encoder and decoder layers, targeting sub-modules such as cross attention, self-attention and feed-forward blocks. 
In attention layers, we also intervene at the head level.

\paragraph{Encoder Lens.}
\label{sec:encoder_lens}

We introduce Encoder Lens, a method for analyzing intermediate encoder representations in ASR models. Inspired by the Diffusion Lens framework for interpreting text encoders in text-to-image models~\cite{toker2024diffusion}, our goal is to examine how representations evolve across encoder layers in encoder–decoder ASR systems.

Given an input audio signal \( \mathbf{x} \), the encoder produces a sequence of hidden vectors \( \mathbf{h}_\tau^{l_e} \in \mathbb{R}^d \) at each layer \( l_e \in \{1, \dots, L_e\} \). For each layer, we extract the full representation \( \mathbf{h}^{l_e} \in \mathbb{R}^{F \times d} \), apply the model's final encoder layer normalization, and pass it directly into the decoder.
As in \citet{toker2024diffusion}, we find that applying the final layer normalization is crucial. Without it, the decoder struggles to produce coherent or grammatical output. 
This process constructs a textual representation for each encoder layer, which we further analyze in Section\ref{sec:encoder_lens}.

\section{Experiments}


Our experiments focus on two state-of-the-art ASR systems with distinct architectural designs:

\textbf{Whisper.}  
We use whisper‑large‑v3 \cite{radford2023robust}, an encoder–decoder model designed for multilingual speech-to-text and speech translation tasks. It features a 32-layer audio encoder and a 32-layer text decoder, trained jointly on large-scale paired audio–text datasets. The model has $\sim1.5$B parameters in total.

\textbf{Qwen2‑Audio.}   
We use Qwen2‑Audio‑7B‑Instruct \cite{chu2024qwen2}, a Large Audio Language Model with $\sim8.2$B parameters. It combines a frozen whisper-large-v3 encoder with a Qwen2.5-7B decoder, trained for multimodal instruction following including audio transcription. The encoder output is prepended to the decoder input as a prefix, enabling the model to handle both spoken and textual instructions.




\subsection{Probing for Transcription Enrichment}
\label{sec:Transcription-Enrichment}

While ASR models are trained to produce transcriptions, both their encoder and decoder layers capture a broad range of information beyond the spoken words. By training simple probes on internal activations (Section~\ref{sec:probing}), we can reveal that specific layers encode various attributes despite these properties not being part of the model's supervision.

Once such attributes are encoded, they remain accessible throughout the forward pass. This means that a single transcription run implicitly generates a much richer representation, capturing both acoustic and contextual information. The examples shown here demonstrate just a few of the many properties that can be extracted from intermediate layers across the model. Full layerwise results and training details appear in the Appendix.

\textbf{Speaker Gender.}
We examine whether speaker gender is encoded in the shared Whisper-large-v3 encoder used by both \whisper{} and \qwen{}. We train linear probes on 2,000 labeled examples from LibriSpeech~\cite{7178964}, and evaluate on 500 samples from the test-clean split. We apply probes to each encoder layer individually. The best performance is observed at layer 25, achieving 94.6\% accuracy, indicating strong linear decodability of gender features in deeper layers.
For comparison, asking Qwen2-Audio to determine speaker gender based on its textual outputs yields only 87.8\% accuracy. This demonstrates that the model knows more than it explicitly shows in its outputs, a phenomenon that was also reported in LLMs \cite{orgad2024llms}, and highlights the advantage of probing internal representations. 



\textbf{Clean vs.\ Noisy Environment.}  
Next, we investigate whether noisy environment is reflected in the Whisper encoder hidden representation. 
We train linear probes using examples from the \texttt{dev-clean} (speech recorded in clean conditions) and \texttt{dev-other} (noisy or challenging conditions) splits of LibriSpeech, and test on the corresponding \texttt{test-clean} and \texttt{test-other} splits. Probes are applied to individual encoder layers. The best performance is observed at layer 27, reaching 90.0\% accuracy, indicating that the encoder effectively separates clean from noisy speech.



 

\textbf{Accents}  
Finally, we assess whether speaker accent is reflected in the Whisper encoder representations by performing multi-class classification over accent categories. Using the English Accent Dataset~\cite{westbrook_english_accent_dataset}, we select four accent groups: New Zealand, Welsh Valleys, South African, and Indian. We train linear probes on 2,400 samples (600 per class), and evaluate on 337 test samples. The best performance is observed at encoder layer 22, reaching 97.0\% average accuracy. Class-wise accuracies are also high: 95.7\% for Indian, 95.8\% for New Zealand, 96.1\% for South African, and 99.2\% for Welsh Valleys. This result suggests that accent information is linearly decodable from intermediate audio representations.



\subsection{Probing for Hallucination Monitoring}
In this section, we investigate whether model hallucinations can be predicted from internal decoder representations. First, we examine whether hallucinations can be predicted from the residual stream. Second, we probe the hidden representations to identify non-speech content, with a focus on hallucinations caused by misinterpreting silence or background noise inputs.

\paragraph{Hallucination Prediction from Decoder Residual Stream.}

Here, we ask whether hallucinations can be predicted in advance by examining the model's internal state.
Inspired by findings in the LLM literature~\citep{oneill2025singledirectiontruthobserver}, we test this hypothesis by linearly probing the ASR decoder's residual stream at the final token position (\texttt{<eos>} token) across all layers.

To that end, we pose a binary classification task to differentiate between samples with zero and high word error rate (WER). We transcribe the test-clean subset of LibriSpeech~\cite{7178964} and CommonVoice 16.1 \cite{ardila2020commonvoicemassivelymultilingualspeech} datasets using each target model, then select 150 samples with zero WER and the 200 samples with highest WER values, creating a 400-sample dataset split 70\%-30\% into training and test sets. This creates a challenging task where each model must distinguish between its own high-quality and severely degraded transcriptions.

We verified that the text length distributions are identical between the low and high WER groups, ensuring that classification performance reflects transcription quality differences rather than length biases.

We train linear probes on the final token's residual stream representation to distinguish between high-quality transcriptions and hallucinations. Surprisingly, the results show that hallucinations can be accurately identified by linear probing the decoder's residual stream, with maximum accuracy of 93.4\% at layer 22. This suggests that \whisper{} encodes quality-related signals deep in the decoder at the completion of text generation, enabling lightweight hallucination prediction directly from internal activations. We repeat the experiment on the Common Voice dataset and achieve 88.1\% accuracy at layer 22, confirming that hallucination-related signals are consistently embedded in the residual stream across domains. 

Next, we conduct the same linear probing experiments with Qwen2-Audio. On the LibriSpeech dataset, the peak accuracy achieved by the linear probe was 70.2\% at layer 22. For the Common Voice dataset, the probe reached an accuracy of 83.6\%, also at layer 22, suggesting consistent architectural patterns for quality encoding across different ASR models. Detailed results across all layers and additional experimental details are provided in the Appendix.

\paragraph{Speech vs. Non-Speech for Non-Speech Hallucinations.} 

Recent studies show that ASR models, such as Whisper, may hallucinate by generating grammatically correct transcriptions for non-speech audio inputs~\citep{baranski2025investigation, frieske2024hallucinations}. 

In this section, we investigate whether internal activations alone can reliably distinguish speech from non-speech inputs, enabling enriched transcript metadata during inference. Such capability would allow systems to flag potentially hallucinated outputs when processing ambiguous audio streams.

To evaluate this, we construct a balanced binary classification dataset of 800 samples: 400 speech samples from LibriSpeech~\cite{7178964}, CommonVoice~\cite{ardila2020commonvoicemassivelymultilingualspeech}, and MLS~\cite{Multilingual_LS}, and 400 non-speech samples from MUSAN~\cite{snyder2015musan}, FSD50K~\cite{fonseca2021fsd50k}, AudioCaps~\cite{kim2019audiocaps}, and generated white noise, encompassing diverse acoustic environments and sound events.
We focus specifically on the more challenging cases where non-speech audios are transcribed into coherent words.

We probe the decoder's hidden states using linear classifiers trained on the final token representation at each layer. The results reveal perfect classification performance (100\% accuracy) from layers 10–28, and near-perfect accuracy (99.17\%) at layer 31. This demonstrates that Whisper encodes speech versus non-speech as a fundamental, linearly separable distinction in its decoder residual stream, despite generating confident transcriptions for both input types.

These findings suggest that trained linear probes could provide real-time speech detection metadata alongside transcriptions, enabling systems to identify and flag potentially hallucinated outputs during inference. See Appendix for detailed dataset construction and full probing results.

\begin{figure}[t]
    \centering
    \includegraphics[width=1.00\linewidth]{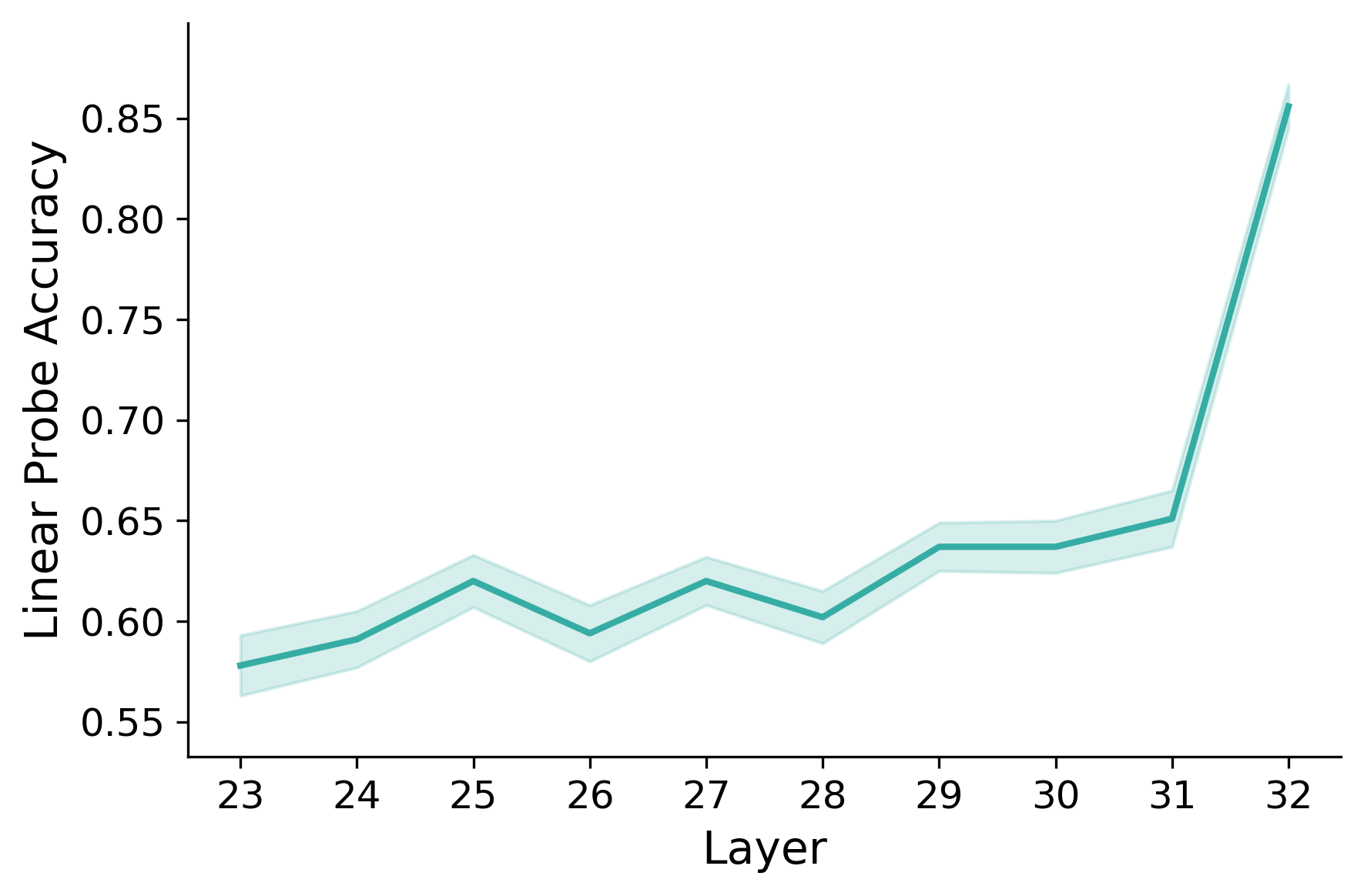}
    \caption{\textit{Whisper encoder understands semantics}: Average linear probe accuracy ($\pm$SEM) across encoder layers for semantic classification.}
    \label{fig:ecoder_semantic}
\end{figure}

\subsection{Analyzing Acoustic, Contextual, and Semantic Mechanisms}

It is well established that the \whisper{} decoder functions as a weak language model~\cite{peng2023reproducing, ellena2023interpreting}, primarily responsible for generating transcriptions based on semantic context rather than purely acoustic cues~\cite{radford2023robust}. The encoder, in contrast, is tasked with capturing the acoustic properties of the input audio~\cite{liu2024exploration}. This apparent division of roles is widely assumed in encoder–decoder ASR systems, yet the extent to which the encoder influences the final transcription output remains largely unexplored.

\paragraph{Acoustic and Contextual Mechanism.}
In the following experiment, we examine whether the model favors the acoustically spoken word, or a more contextually plausible alternative. To investigate the acoustic–contextual mechanism tradeoff, we construct a dataset of synthetic audio samples generated using a text to speech model.

Each sentence is designed to trigger contextual errors in the model, containing an acoustically ambiguous word: the true spoken word is atypical or contextually unexpected, while a more plausible word with similar phonetics fits the surrounding context. For example, a speaker may say \emph{``white lice''} (acoustic truth) in a context where \emph{``white rice''} would be expected. The constructed dataset consists of 700 such examples in total. The Whisper model made contextual errors on 153 examples, which we analyze to understand the underlying mechanism. Qwen2-Audio produced errors on 251 instances, indicating a stronger tendency toward contextual predictions compared to Whisper. In both cases, the model's output differs from the ground truth by a single target word, enabling precise analysis of the acoustic–contextual tradeoff.

Next, we perform \textit{component patching} across all encoder and decoder subcomponents on the 251 Qwen2-Audio cases and 153 Whisper cases. Motivated by established intervention methods \ref{sec:patching_methods}, we select a white noise audio as the disruptive audio, $\mathbf{x}^{\mathrm{dis}}$ (see Section~\ref{sec:methods}).

Surprisingly, applying patching interventions to encoder components using disruptive audio improves acoustic accuracy, despite the common assumption that encoders operate purely on acoustic input, without encoding contextual or semantic information (Table~\ref{tab:acoustic_contextual_patching}). Both encoder and decoder components contributed to restoring acoustic accuracy, with encoders showing particularly strong effectiveness across both models.

These findings indicate that the encoder is not limited to acoustic processing, it also encodes contextual and semantic expectations that can bias the model toward more likely completions. In fact, intervening on the encoder improves transcription accuracy in many cases, providing direct evidence that semantic influence originates in the encoder and that not all contextual decisions are made in the decoder.

\begin{table}[t]
\centering
\begin{tabular}{lcc}
\toprule
\textbf{Metric} & \textbf{Whisper} & \textbf{Qwen2-Audio} \\
\midrule
Error Cases & 153/700 (21.8\%)& 251/700 (35.8\%)\\
Restored Acc. & 136/153 (88.9\%)& 176/251 (70.1\%)\\
Via Encoder & 130/153 (85.0\%)& 171/251 (68.1\%)\\
Via Decoder & 126/153 (82.4\%)& 147/251 (58.6\%)\\
\bottomrule
\end{tabular}
\caption{Acoustic-contextual patching results using white noise disruption.}
\label{tab:acoustic_contextual_patching}
\end{table}

\paragraph{Whisper Encoder Understands Semantics.}
Following our findings that disrupting encoder components paradoxically improves acoustic transcription, here, we aim at investigating whether ASR encoders encode semantic information.
To that end, we design and construct a dedicated synthetic audio dataset.
The dataset consists of carefully selected terms from distinct semantic groups, e.g., fruits and clothing. Next, we train linear probes to discriminate between pairs of these semantic categories based solely on encoder activations.

\begin{figure}[h!]
\centering
\includegraphics[width=1.0\columnwidth]{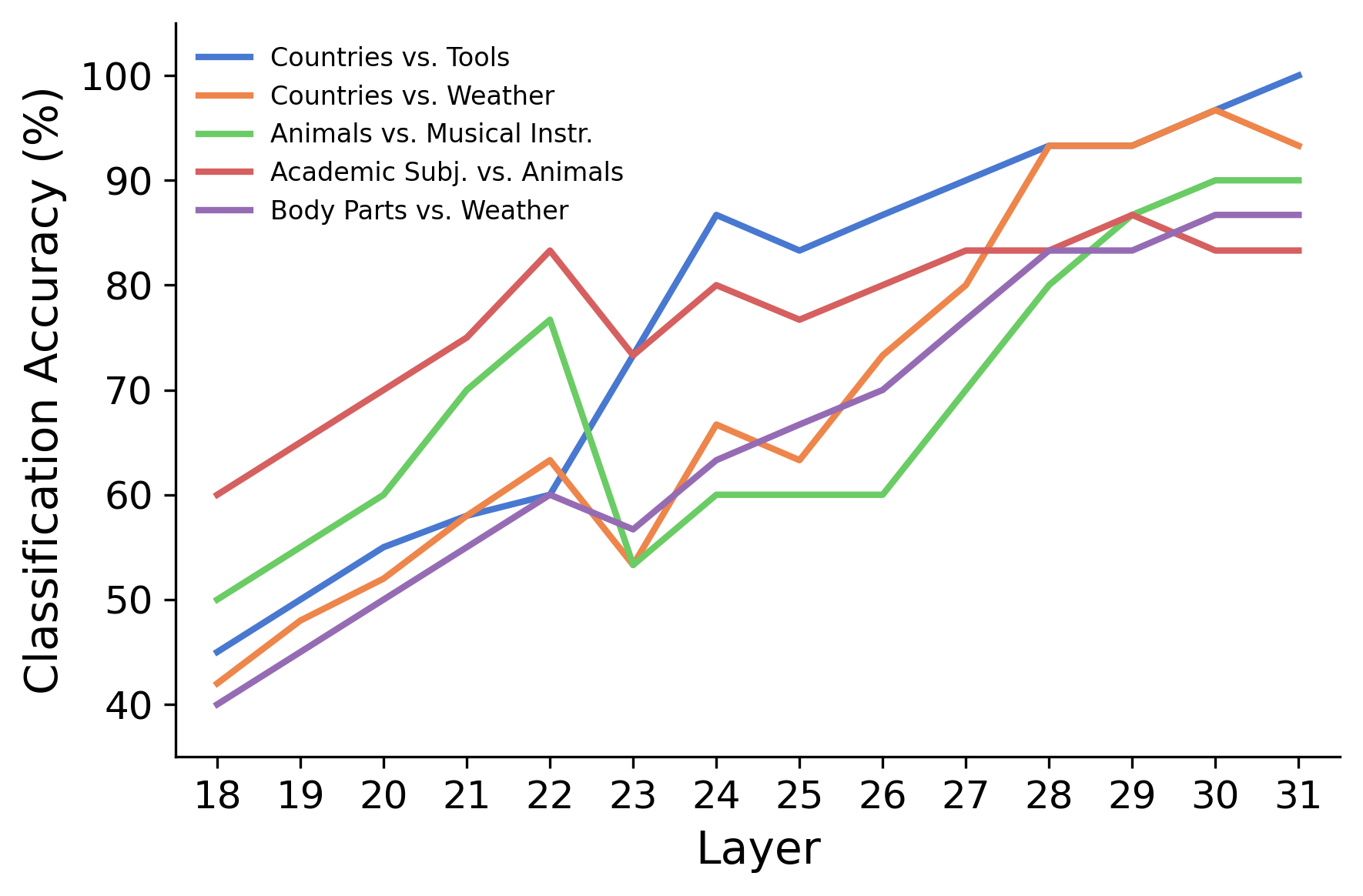}
\caption{Semantic classification progression across encoder layers for selected category pairs.}
\label{fig:semantic_progression}
\end{figure}

The probe results demonstrated a clear pattern: semantic encoding becomes increasingly prominent in higher encoder layers. Figure~\ref{fig:ecoder_semantic} shows the mean classification accuracy across 66 semantic category pairs, with many individual pairs achieving exceptional performance. Remarkably, Figure~\ref{fig:semantic_progression} reveals that semantic understanding emerges as early as the middle encoder layers (18-21), with several category pairs already achieving substantial performance well before the final layers. This early semantic emergence demonstrates a gradual build-up of semantic representations throughout the encoder hierarchy.

In the last encoder layer, probes reached their peak semantic understanding, with average accuracy of 85.6\%. Several category pairs achieving 96.7\% accuracy (e.g., distinguishing \emph{countries} from \emph{weather} or \emph{clothing}), while maintaining strong performance across most semantic distinctions. The progression from early semantic signals to sophisticated categorical distinctions suggests that the encoder develops hierarchical semantic representations alongside its acoustic processing capabilities.
Full evaluation details in the Appendix.



\begin{figure*}[t]
    \centering
    \subcaptionbox{Selected Token Probabilities\label{fig:token-select}}[0.32\linewidth]{
        \includegraphics[width=\linewidth]{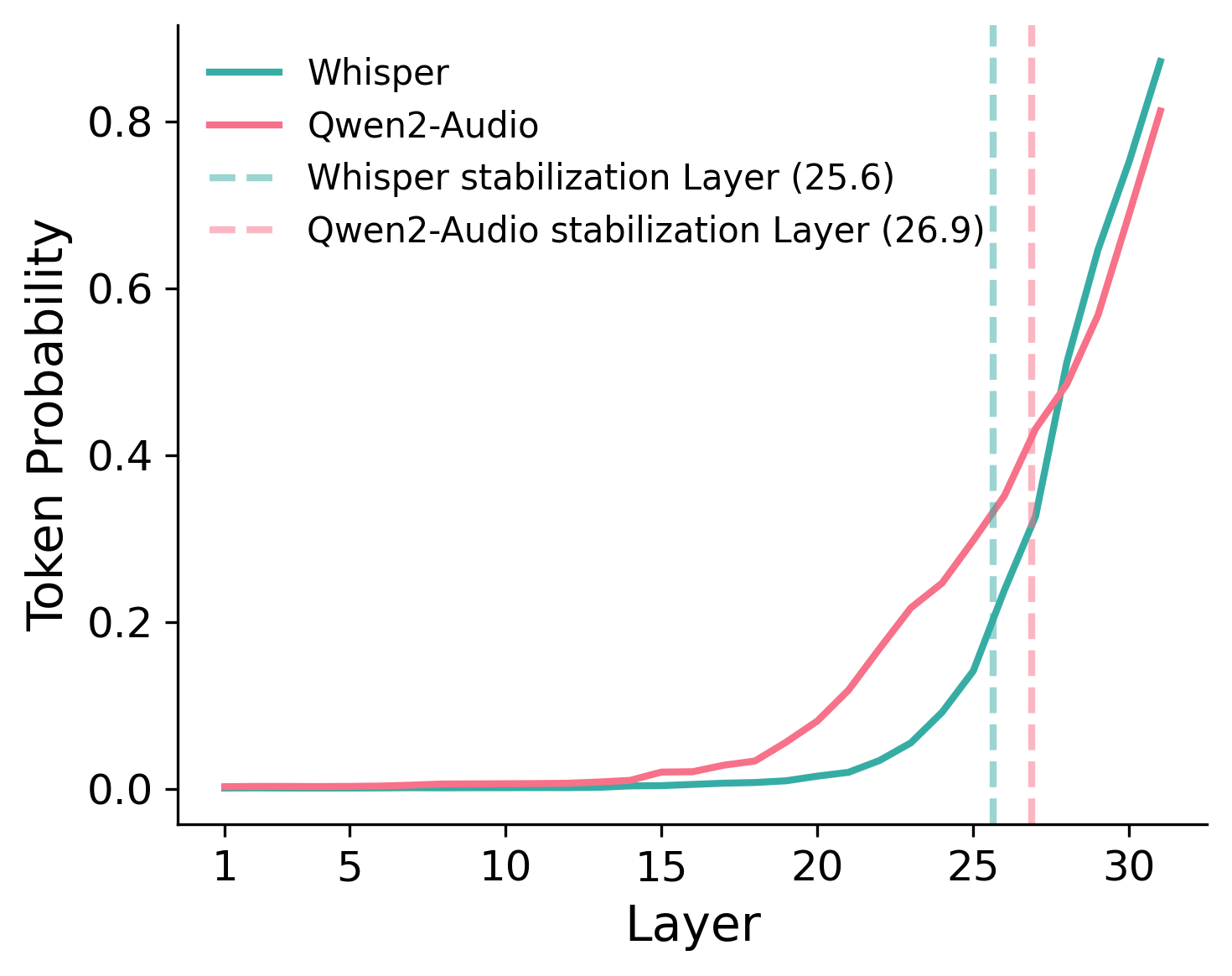}
    }
    \hfill
    \subcaptionbox{Selected Token PER\label{fig:per}}[0.32\linewidth]{
        \includegraphics[width=\linewidth]{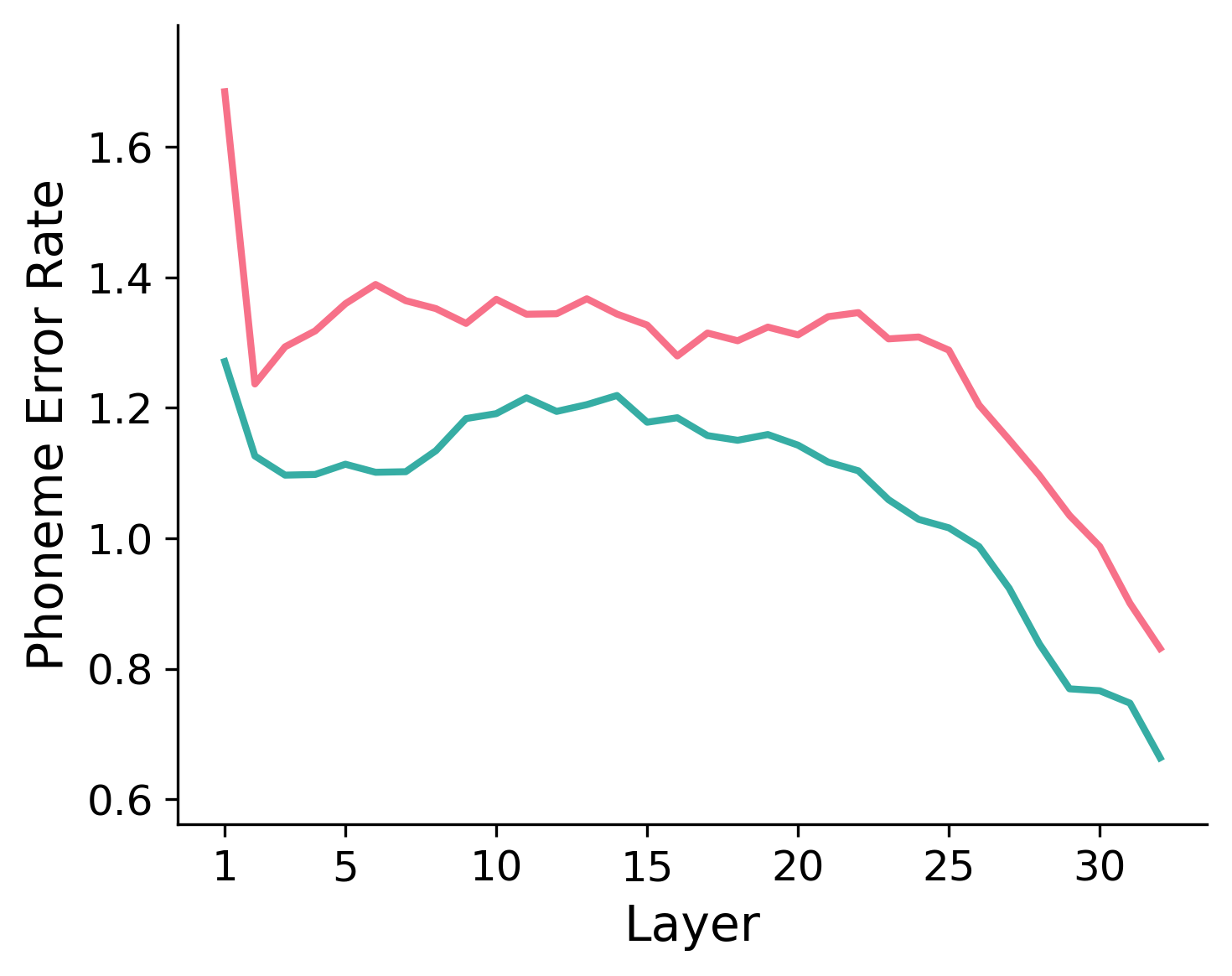}
    }
    \hfill
    \subcaptionbox{Selected Token Cos-Sim\label{fig:cos-sim}}[0.32\linewidth]{
        \includegraphics[width=\linewidth]{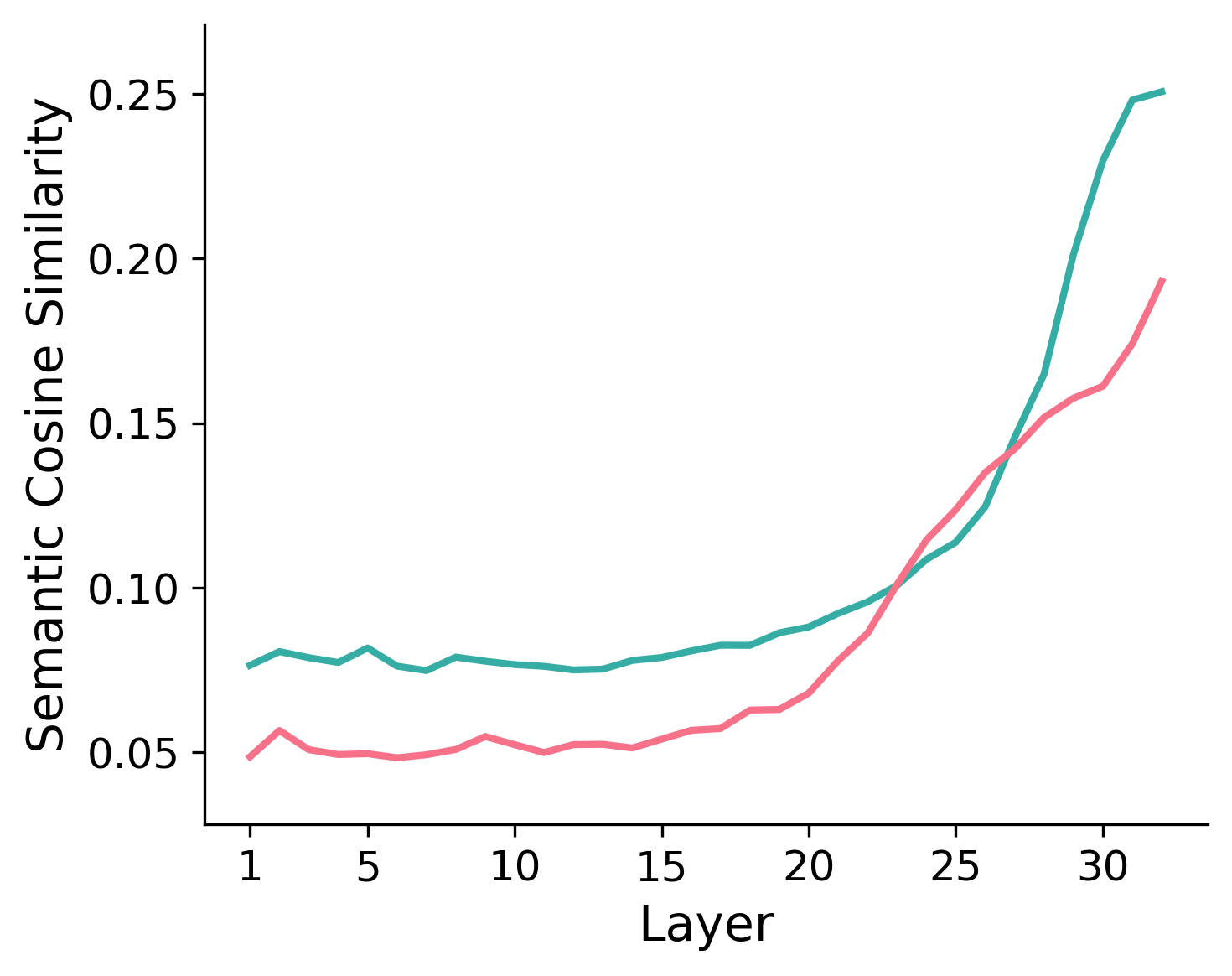}
    }
    \caption{\textit{Token Selection Mechanism}: (a) Probability of the final selected token across decoder layers, with indication to the average saturation layer; (b) Phoneme error rate (PER) by layer. Lower PER indicates higher acoustic similarity; (c) Tokens semantic cosine similarity by layer. Higher similarity indicates greater semantic similarity between top-5 tokens.}
    \label{fig:main}
\end{figure*}

\subsection{Token Selection Mechanism}
In this section, we explore the internal mechanism that underlies token selection within the decoder. Our aim is to understand when the model determines which tokens to output. To this end, we apply the \textit{logit lens} technique to analyze model's behavior. Specifically, we evaluate both \whisper{} and \qwen{} in six languages: English, French, Spanish, German, Chinese and Italian. For each language, we randomly sample 100 utterances from the CommonVoice \cite{ardila2020commonvoicemassivelymultilingualspeech} test set, resulting in a balanced multilingual evaluation set.

\paragraph{Token Selection Dynamics.} We start by examining how the probability assigned to the final selected token changes across different layers. Figure \ref{fig:token-select} presents the mean probability averaged across examples from all the six languages and token positions.
For both \whisper{} and \qwen{}, the probability remains low until around layer 20, after which it increases sharply, with the final three layers showing high confidence in the selected token.
We also analyze the saturation layer \cite{lioubashevski2024looking}, defined in Section \ref{sec:methods}. Interestingly, although the mean probability of the selected token is generally higher in \qwen{}, the saturation layer tends to appear earlier in \whisper{}. We provide additional results of the token selection by language in the Appendix.

\paragraph{Acoustic and Semantic Token Similarity.} 
Both \whisper{} and \qwen{} generate transcripts as sequences of subword tokens, in contrast to models such as \cite{baevski2020wav2vec} that operate at the grapheme level. Given that tokens can differ in phonetic and acoustic content, we extend our analysis by comparing the acoustic distance and semantic similarity between the final selected token and the top five candidate tokens produced by the model at each layer.
For acoustic distance, we use a modified version of the Phoneme Error Rate (PER) metric. As with the standard PER, lower values indicate higher acoustic similarity. For semantic similarity, we compute cosine similarity between token embeddings. Details on how these metrics are calculated appear in the Appendix.

Figure~\ref{fig:per} shows that across all layers, \whisper{} consistently achieves lower PER scores than \qwen{}, suggesting higher acoustic similarity to the final selected token. Both models also display a notable PER drop around layer 25, aligning with the saturation point where predictions stabilize. This suggests that from layer 25 onward, not only does the model converge on the final prediction, but the other top-5 candidate tokens also share closer acoustic characteristics with it.

While one might expect \qwen{} to outperform in semantic similarity given its large language modeling capacity, Figure~\ref{fig:cos-sim} reveals that \whisper{} actually maintains higher semantic similarity scores across most layers, indicating its top candidate tokens remain more semantically aligned with the final output.

\paragraph{Next-Token Prediction Capabilities.} Finally, we examine the model's ability to anticipate future tokens, i.e., tokens of future timestamps, beyond the current selection in step $s$. We observe that \qwen{} begins ranking the immediate next token ($s+1$) among its top candidates around layer 21 and retains some predictive ability for the token at position $s+2$. In contrast, \whisper{} shows a later but sharper improvement, with notable gains starting around layer 29. Full details are provided in the Appendix.

\subsection{Decoder Repetition Mechanisms in Whisper}

Repetition hallucinations, where decoder-based models produce excessively repetitive output, are a well-documented failure mode across both language and speech domains \cite{baranski2025investigation, yona2025interpreting}. 

Whisper occasionally produces repetitive outputs \cite{baranski2025investigation}. Based on our observations, these phenomena occur in several specific scenarios: when the input audio itself contains repetitive content (e.g., saying "hey" ten times results in Whisper generating hundreds of repetitions), during code-switching between languages, and when processing fragmented, heavily noised, or completely unclear audio inputs (see examples in Appendix).

We hypothesized that such hallucinations stem from specific components within the decoder's attention mechanisms, rather than being the result of a distributed failure across the model. To test this, we apply both causal patching and ablation interventions (See section \ref{sec:patching_methods}) on the Whisper model. For each of the decoder's 32 layers, we modified the outputs of three core components: cross-attention,
self-attention, and feed-forward layers. Patching involved replacing internal activations with those from clean, non-repetitive reference audio, while ablation zeroed out the original activations. For evaluation, we construct a multilingual dataset of 102 utterances prone to repetition hallucinations, sampled from the Japanese and English portions of CommonVoice 16.1.
\cite{ardila2020commonvoicemassivelymultilingualspeech}.

Our results show that intervening on cross-attention substantially reduced repetitions. Patching in layer 23 resolved 76\% of cases, and layer 18 covered an additional 13\%. Self-attention and fully-connected interventions had no measurable effect. Moreover, a head-level analysis revealed that head 13 in layer 18 was especially influential, suppressing repetition by 78.1\% when targeted alone. This single-head effectiveness represents a remarkably focused intervention: out of 640 total attention heads in the model (32 layers × 20 heads), one specific head in the cross-attention mechanism appears to play a disproportionately critical role in repetition control. Combined, layer 23 and head 13 in layer 18 accounted for 89\% of the corrected examples.

The concentration of repetition control in specific cross-attention components reveals these hallucinations are highly localized. The remarkable effectiveness of a single attention head demonstrates significant progress toward mechanistic understanding and enables targeted interventions - these components can be monitored, steered, or fine-tuned to suppress errors without degrading core performance.

\subsection{Encoder Lens}\label{EncoderLens}

In this experiment, we aim to gain a deeper understanding of how representations evolve across the encoder layers. To this end, we use the \textit{encoder lens} technique described in Section~\ref{sec:encoder_lens}, which involves omitting the top layers of the encoder and directly passing intermediate representations to the decoder. 
We analyzed $400$ audio samples for both Whisper and \qwen{}, drawn from English (LibriSpeech, \citet{7178964}), Spanish (Multilingual LibriSpeech, \citet{Multilingual_LS}), and Chinese (AISHELL, \citet{aishell_2017}). These samples were randomly selected to ensure typological and phonetic diversity across languages and datasets.

The results show that Whisper exhibits a highly structured representational hierarchy. Layers 0 to 22 mainly produce empty strings or isolated punctuations. At the later layers, the model sometimes produces short, often incomplete, words or monosyllabic tokens, that sometimes match the beginning of the actual utterance. We observe a recurring phenomenon at the 20th layer and up to the 27th: the model occasionally outputs syntactically well-formed phrases. The start of the phrases resemble to the start of the audio content, followed with unrelated text. This text, however, is grammatically correct. For example, in one of the samples, the full correct phrase is:
\texttt{ Yes, I need repose. Many things have agitated me today, both in mind and body. When you return tomorrow, I shall no longer be the same man.}
and the output of the 26 layer is:
\texttt{ Yes, I need to go to the bathroom.}
Which is grammatically coherent but does not match the content in the original audio. This suggests that in this mid-layer zone, Whisper may begin to behave like a loosely grounded language model, producing fluent but unanchored completions~\cite{chen2024language}. 
Another interesting pattern we observed begins at the 27th encoder layer, where the model starts to fall into repetition loops. This phenomenon is consistent across all languages. This behavior intensifies through the 30th layer, which appears to be the most consistently affected. In our Whisper analysis, around 60\% of the samples showed this repetition pattern. Only in the final layers (31st and 32nd) these repetitions resolve into fluent, grammatically correct transcriptions.

\qwen{}, in contrast, reveals different failure patterns. While the last five layers reliably generate accurate transcriptions, earlier layers show severe degradation. We perform a frequency analysis of the \qwen{} results, which reveals a surprising phenomenon: the phrase \texttt{Kids are talking by the door} (potentially from the RAVDESS \cite{livingstone2018ryerson} dataset for emotion detection) appears at least once in \emph{390 out of 400 files}, regardless of the input language. This phenomenon is signaling a strongly memorized training data in the model. Alongside it, several high-frequency Chinese expressions which roughly translates to \texttt{Aren't you bored being alone?}) dominate the output in the earlier layers. We provide additional examples in the Appendix. 

These patterns suggest that the model reverts to memorized sequences when uncertain, possibly due to training data imbalance.









\section{Discussion}

This work provides a first comprehensive exploration of interpretability in modern ASR models. We show that a range of acoustic, semantic, and contextual factors are internally represented and can be analyzed using adapted techniques from LLM interpretability. Our experiments uncover localized mechanisms behind hallucinations, repetition loops, and context-driven errors, and offer new tools for inspecting how predictions evolve across layers. 

We demonstrate the potential of interpretability for advancing diverse future research in ASR.These include building internal monitors for hallucination or saturation, developing fine-grained editing and debugging tools for ASR, and informing architectural choices that better balance grounding and fluency. The ability to trace errors back to individual components may also enable targeted interventions or model compression strategies. 




\bibliography{aaai2026}

\clearpage
\newpage

\onecolumn
\appendix

\twocolumn[
\begin{center}
{\Large \bf Appendix: Beyond Transcription—Mechanistic Interpretability in ASR\par}
\vspace{0.5em}
\end{center}
]


\section{Detailed Probing Results}

This section presents comprehensive layer-wise analysis of various probing tasks across Whisper's encoder and decoder layers, demonstrating the model's ability to linearly separate different types of information at different depths.

\subsection{Gender classification}

\begin{figure}[t]
\centering
\includegraphics[width=0.85\columnwidth]{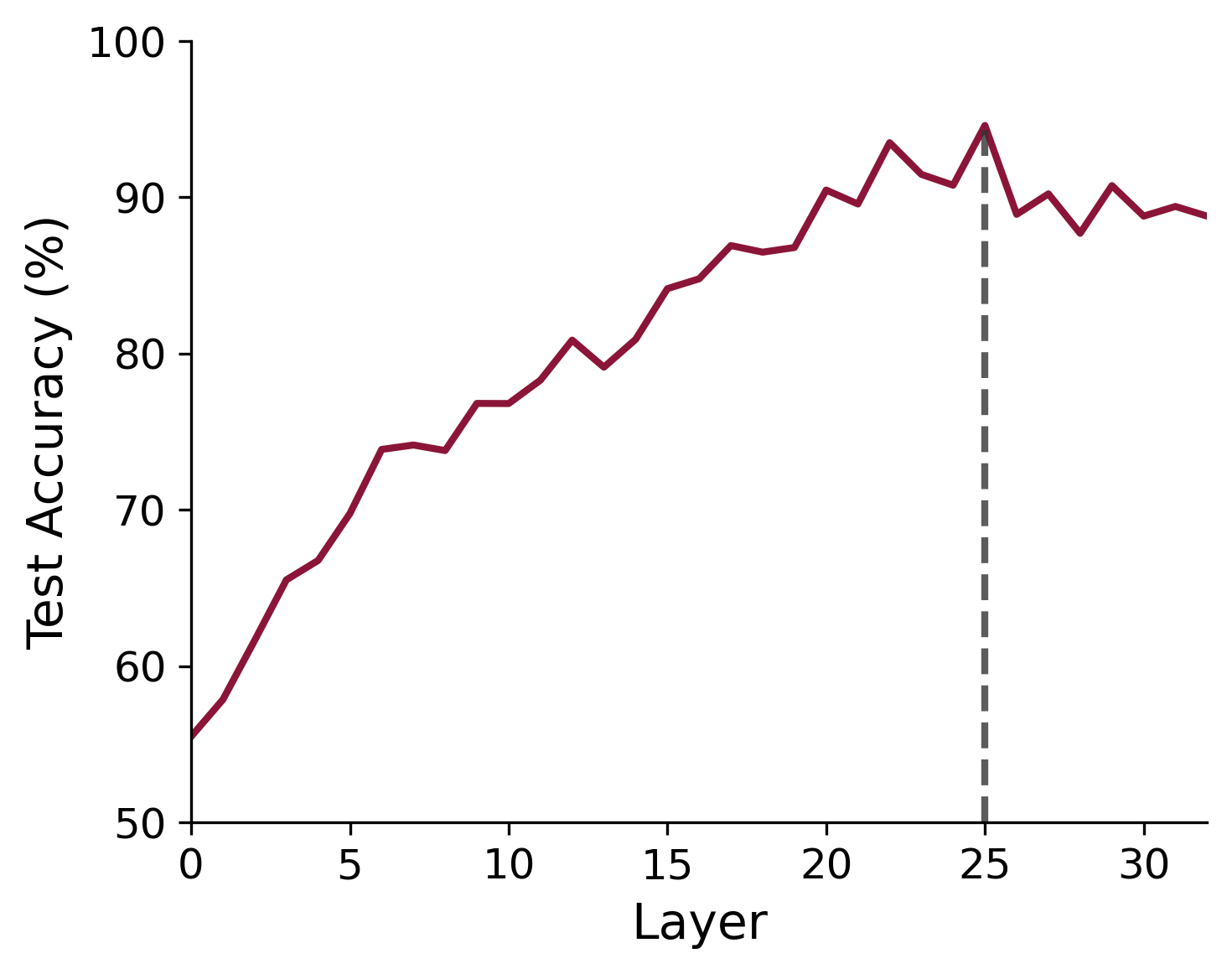}
\caption{Gender classification test accuracy across Whisper encoder layers. Peak performance of 94.6\% achieved at layer 25.}
\label{fig:gender_accuracy}
\end{figure}

Gender classification performance across encoder layers is shown in Figure~\ref{fig:gender_accuracy}. The results demonstrate a clear progression from random performance at the embedding layer (Layer 0: 55.5\%) to peak performance at Layer 25 (94.6\%), indicating that gender information becomes increasingly linearly separable in deeper encoder layers.

\subsection{Clean vs. noisy environment classification}
\begin{figure}[t]
\centering
\includegraphics[width=0.85\columnwidth]{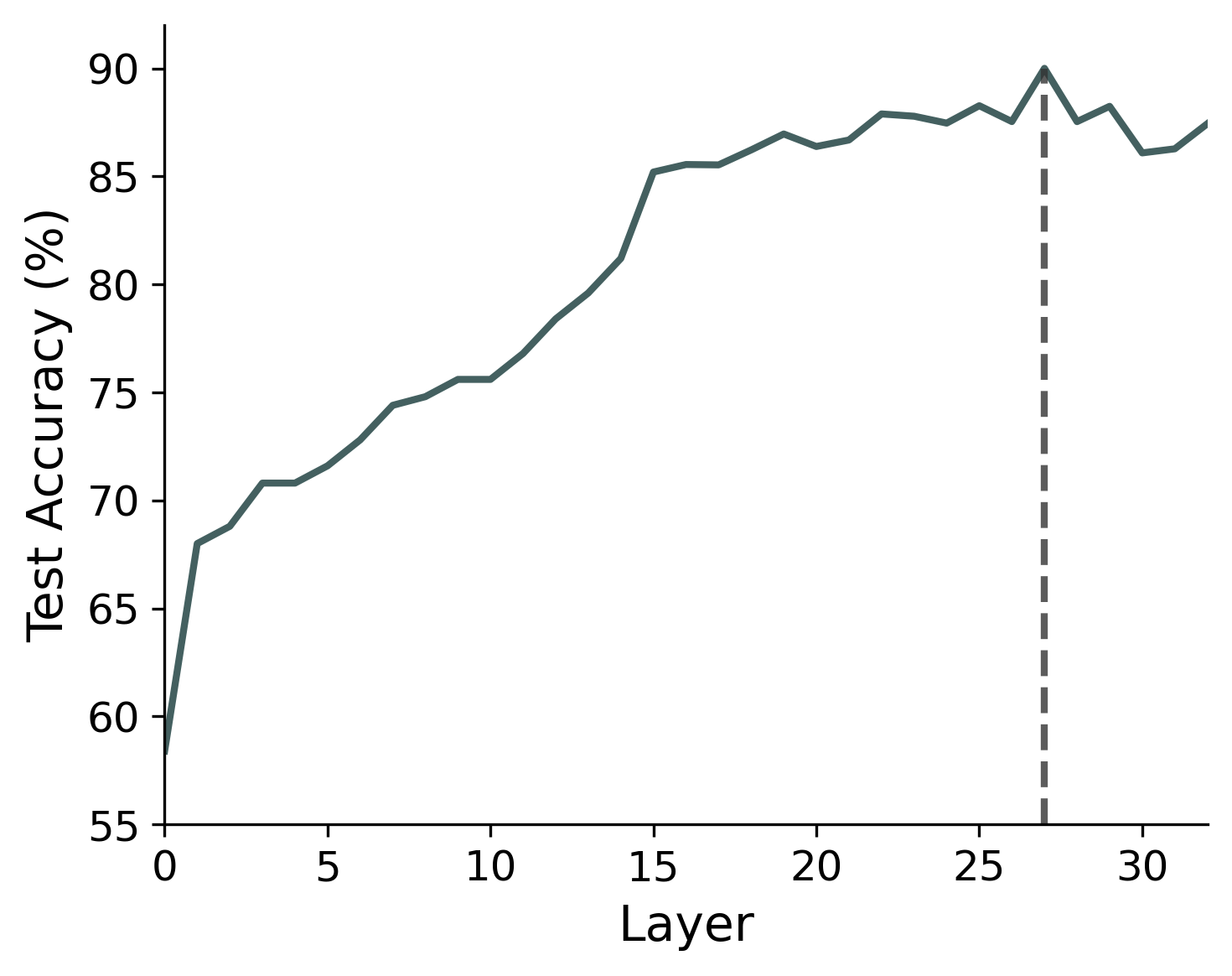}
\caption{Clean vs noisy speech classification accuracy across Whisper encoder layers. Peak performance of 90.0\% achieved at layer 27.}
\label{fig:clean_noisy_accuracy}
\end{figure}

Clean vs. noisy environment classification across encoder layers is presented in Figure~\ref{fig:clean_noisy_accuracy}. The model shows steady improvement from Layer 0 (58.4\%) reaching optimal performance at Layer 27 (90.0\%), suggesting that acoustic quality indicators are most effectively captured in the deeper encoder representations.

\subsection{Accent classification environment classification}
Detailed accent classification results across encoder layers are provided in Table~\ref{tab:accent_layer_results}. The analysis covers four accent categories (New Zealand, Welsh, South African, and Indian) with peak overall accuracy achieved at Layer 22.

\subsection{Speech vs. Non-Speech Classification}

\begin{figure}[h]
\centering
\includegraphics[width=0.85\columnwidth]{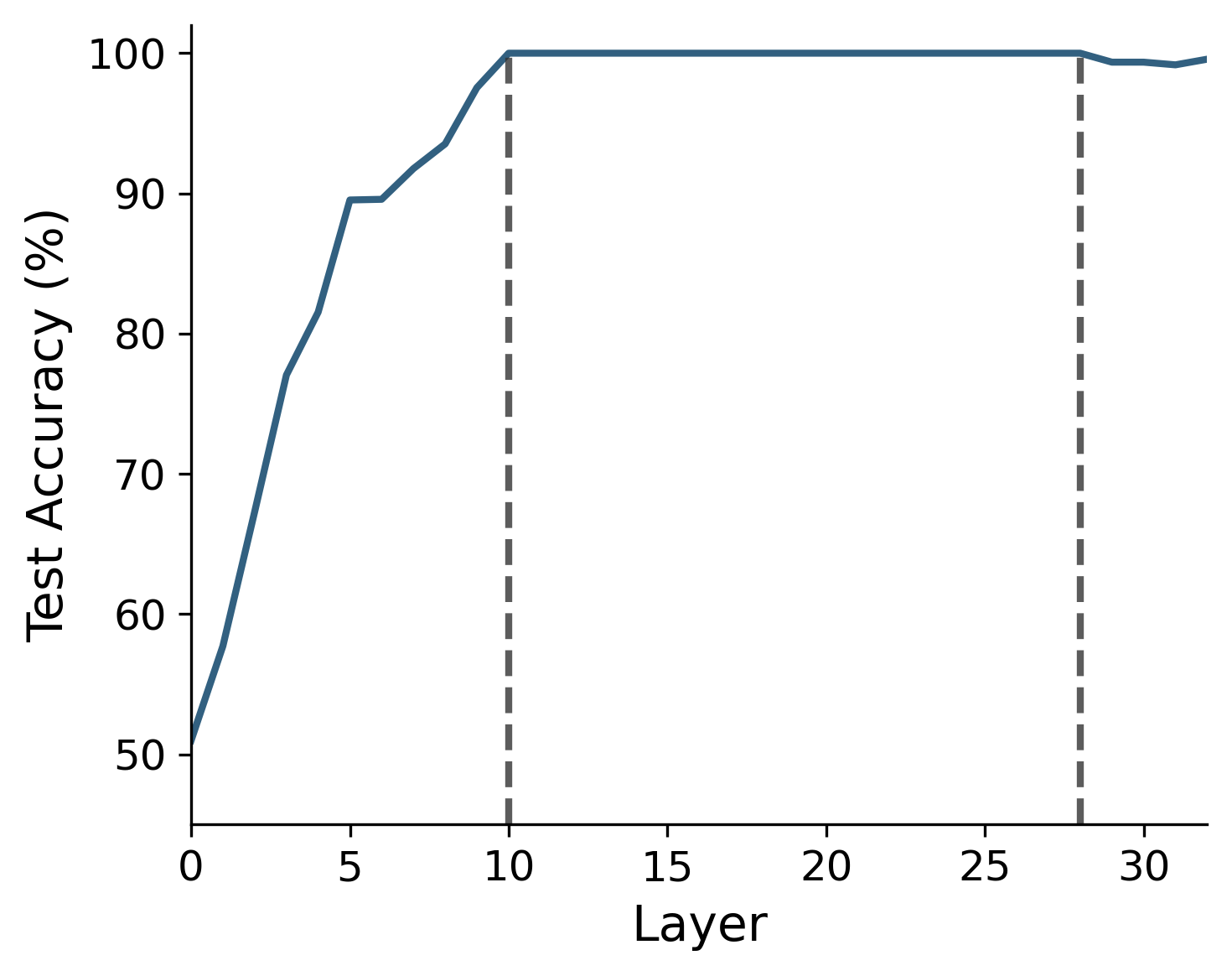}
\caption{Speech vs non-speech classification accuracy across Whisper decoder layers. Perfect classification (100\%) achieved in layers 10-28.}
\label{fig:speech_nonspeech_accuracy}
\end{figure}

Speech vs. non-speech classification performance across decoder layers is illustrated in Figure~\ref{fig:speech_nonspeech_accuracy}. The results reveal remarkable performance, with perfect classification (100\%) consistently achieved across Layers 10-28, and near-perfect accuracy (99.17\%) maintained at Layer 31. This demonstrates the decoder's robust ability to distinguish between genuine speech content and non-speech audio inputs.

\subsection{Hallucination Prediction rom Decoder Residual Stream}

Word Error Rate (WER) classification results for detecting transcription quality are detailed in Table~\ref{tab:librispeech_decoder_detailed}. The decoder residual stream analysis shows peak performance at Layer 22 (93.3\% accuracy), indicating that hallucination-related signals are effectively encoded in the deeper decoder layers during generation.

\begin{table}[t]
\centering
\caption{Accent Classification by Encoder Layer}
\scriptsize
\label{tab:accent_layer_results}
\begin{tabular}{cccccc}
\toprule
\textbf{Layer} & \textbf{Overall} & \textbf{NZ} & \textbf{Welsh} & \textbf{SA} & \textbf{Indian} \\
\midrule
5 & 59.5\% & 38.0\% & 70.0\% & 70.0\% & 60.0\% \\
10 & 63.0\% & 48.0\% & 70.0\% & 66.0\% & 68.0\% \\
15 & 64.5\% & 56.0\% & 76.0\% & 66.0\% & 60.0\% \\
20 & 79.0\% & 74.0\% & 94.0\% & 74.0\% & 74.0\% \\
25 & 82.5\% & 84.0\% & 98.0\% & 68.0\% & 80.0\% \\
28 & 80.5\% & 88.0\% & 96.0\% & 60.0\% & 78.0\% \\
31 & 78.5\% & 82.0\% & 96.0\% & 64.0\% & 72.0\% \\
\bottomrule
\label{tab:accent}
\end{tabular}
\end{table}

\begin{table*}[h]
\centering
\caption{Hallucination Prediction from Decoder Residual Stream }
\label{tab:librispeech_decoder_selected}
\small
\begin{tabular}{cccccc}
\toprule
\textbf{Layer} & \textbf{Test Acc} & \textbf{Train Acc} & \textbf{Train Time (s)} & \textbf{Low WER F1} & \textbf{High WER F1} \\
\midrule
5 & 0.622 & 0.571 & 0.92 & 0.393 & 0.726 \\
10 & 0.900 & 0.838 & 0.76 & 0.907 & 0.892 \\
15 & 0.889 & 1.000 & 0.70 & 0.886 & 0.891 \\
20 & 0.878 & 1.000 & 0.66 & 0.874 & 0.882 \\
22 & \textbf{0.933} & 1.000 & 0.66 & 0.933 & 0.933 \\
25 & 0.900 & 0.995 & 0.64 & 0.899 & 0.901 \\
28 & 0.922 & 1.000 & 0.59 & 0.923 & 0.921 \\
31 & 0.932 & 1.000 & 0.57 & 0.935 & 0.932 \\
\bottomrule
\label{tab:librispeech_decoder_detailed}
\end{tabular}
\end{table*}

\begin{figure}[h]
    \centering
    \begin{subfigure}{0.8\linewidth}
        \centering
        \includegraphics[width=\linewidth]{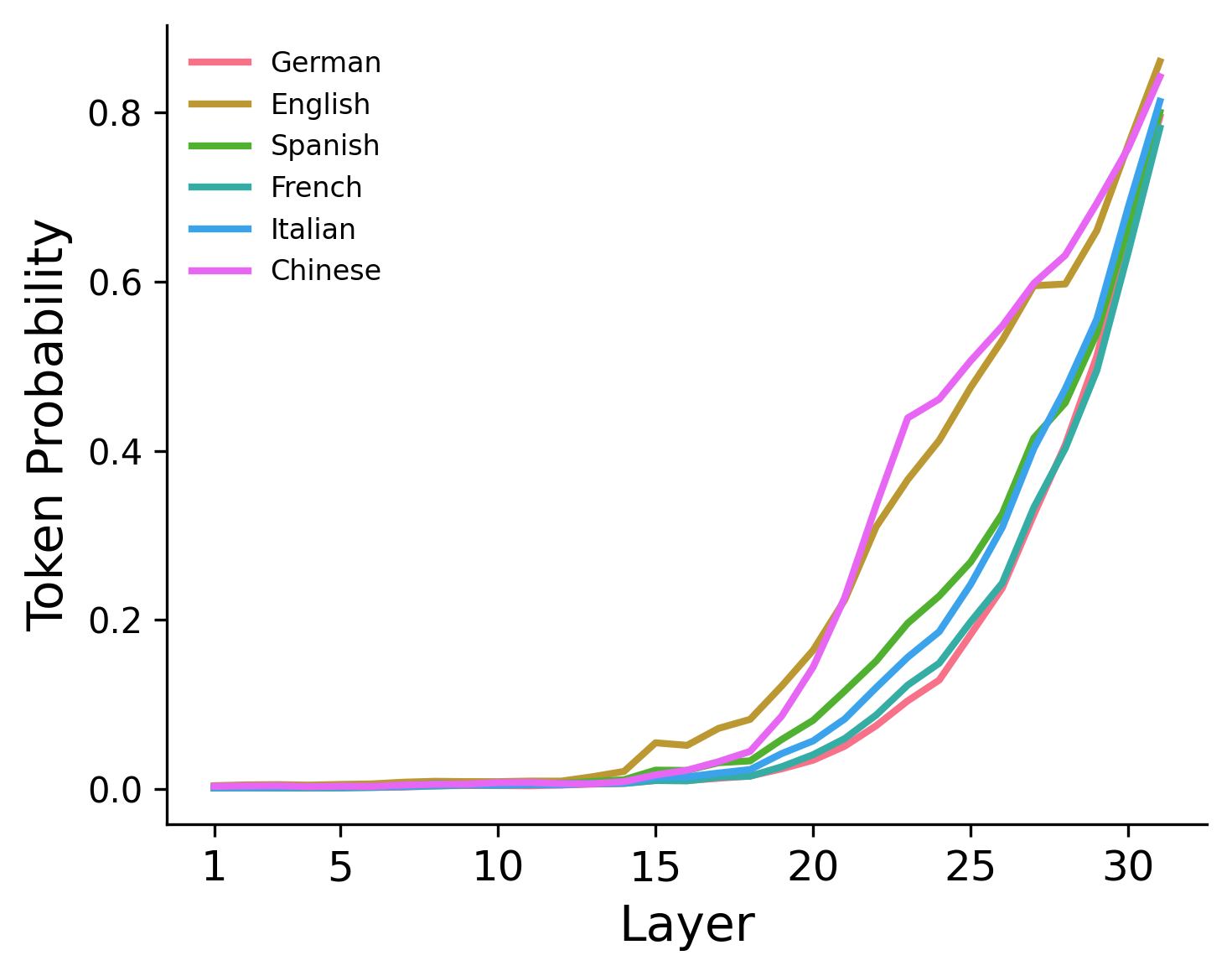}
        \caption{\qwen{}}
        \label{fig:qwen_prob}
    \end{subfigure}
    \vspace{0.5em}
    \begin{subfigure}{0.8\linewidth}
        \centering
        \includegraphics[width=\linewidth]{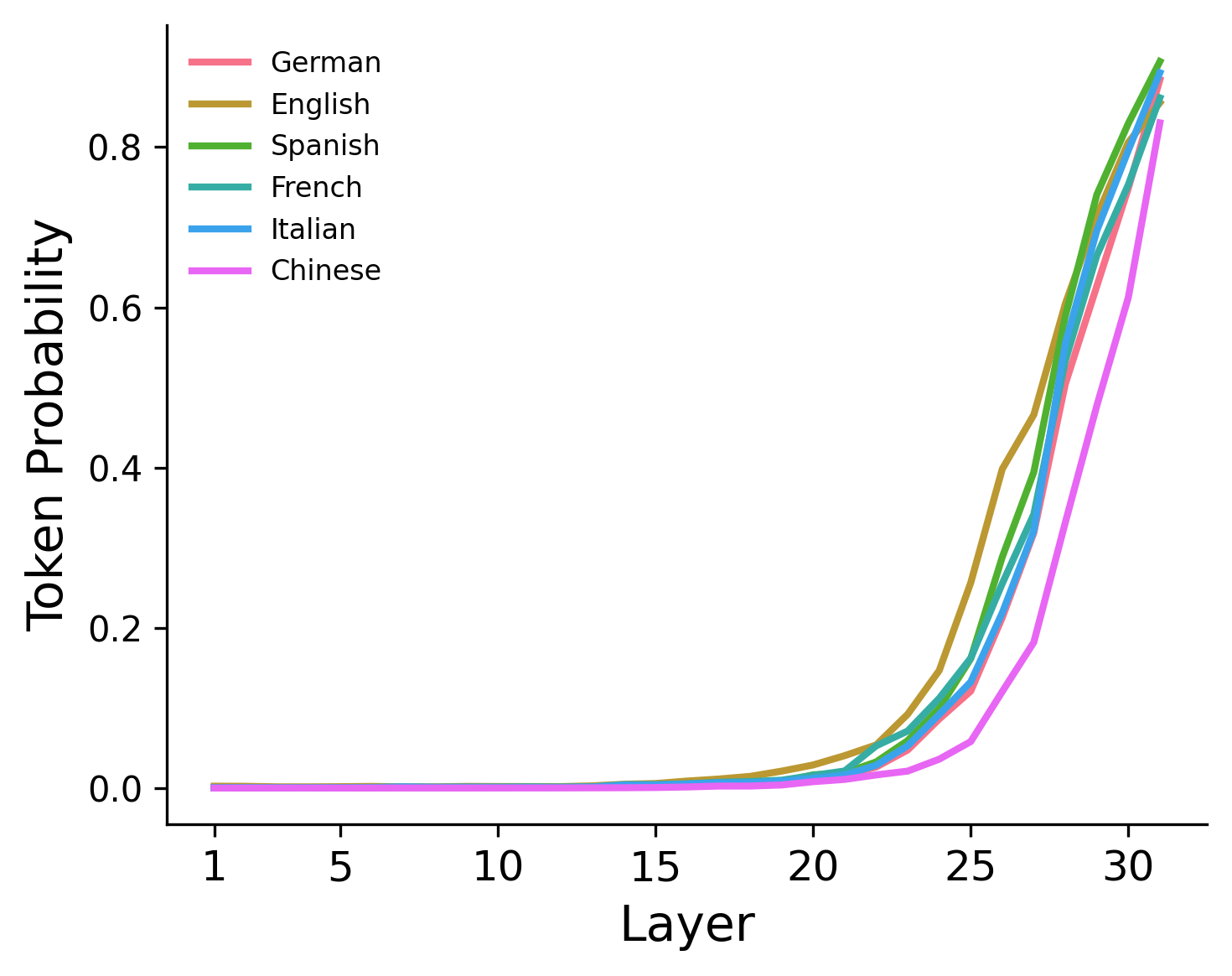}
        \caption{\whisper{}}
        \label{fig:whisper_prob}
    \end{subfigure}
    \caption{Probability of the final selected token across decoder layers and languages for (a)~\qwen{} and (b)~\whisper{}.}
    \label{fig:logit_lens_prob_lang}
\end{figure}


\section{Token Selection Mechanism}
\label{app:logit_lens}

\paragraph{Selected Token Dynamic.}

Figure~\ref{fig:logit_lens_prob_lang} shows the mean probability of the final selected token across layers, broken down by language. The \whisper{} model displays similar trends across all six languages, with slightly higher confidence for English and the lowest confidence for Chinese. In contrast, \qwen{} exhibits a different pattern: for English and Chinese, it shows much higher confidence in earlier layers compared to the other languages.

\paragraph{Top-K Tokens Dynamic.}


\begin{figure}[h]
    \centering
    \begin{subfigure}{0.8\linewidth}
        \centering
        \includegraphics[width=\linewidth]{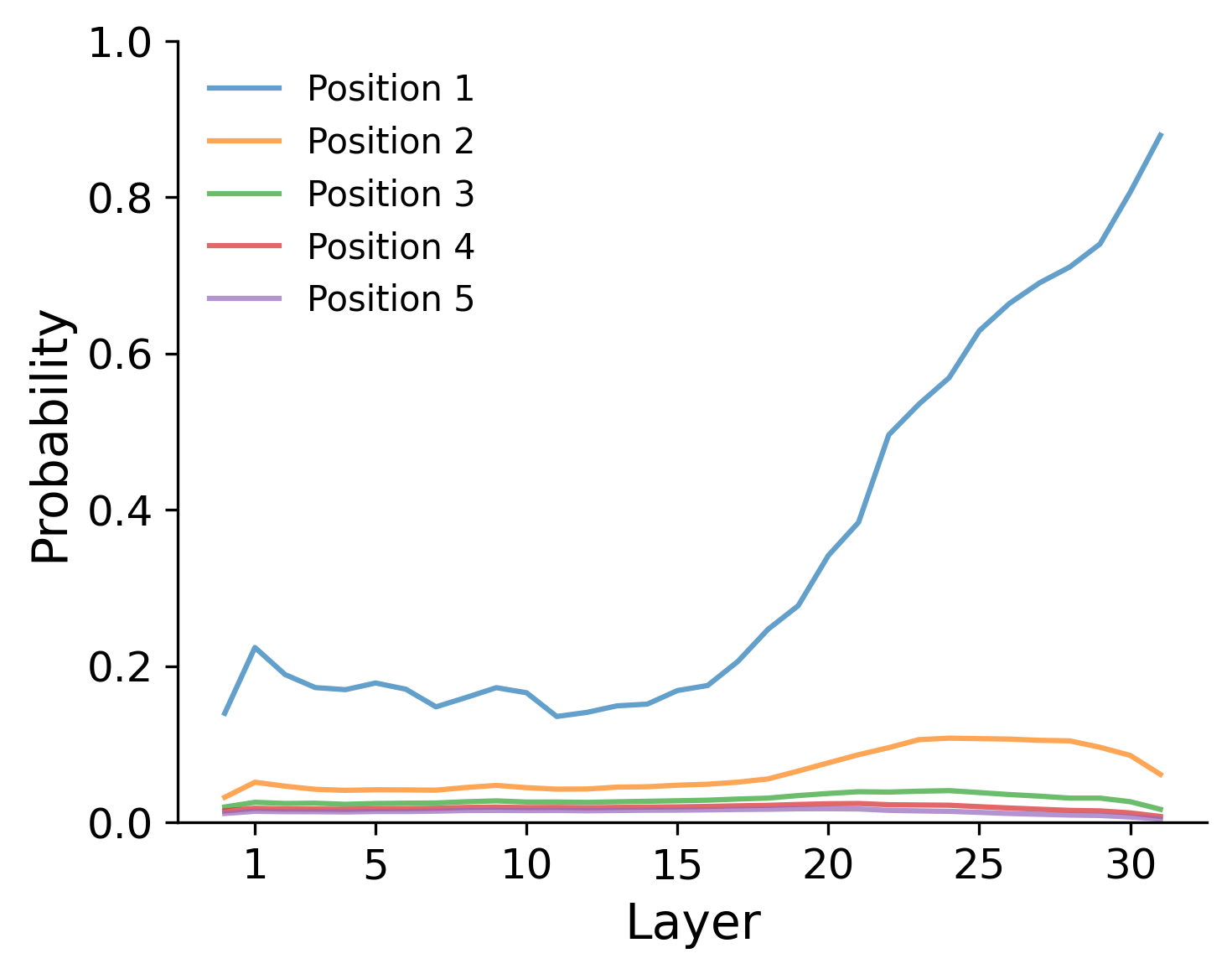}
        \caption{\qwen{}}
        \label{fig:qwen_topk}
    \end{subfigure}
    \vspace{0.5em}
    \begin{subfigure}{0.8\linewidth}
        \centering
        \includegraphics[width=\linewidth]{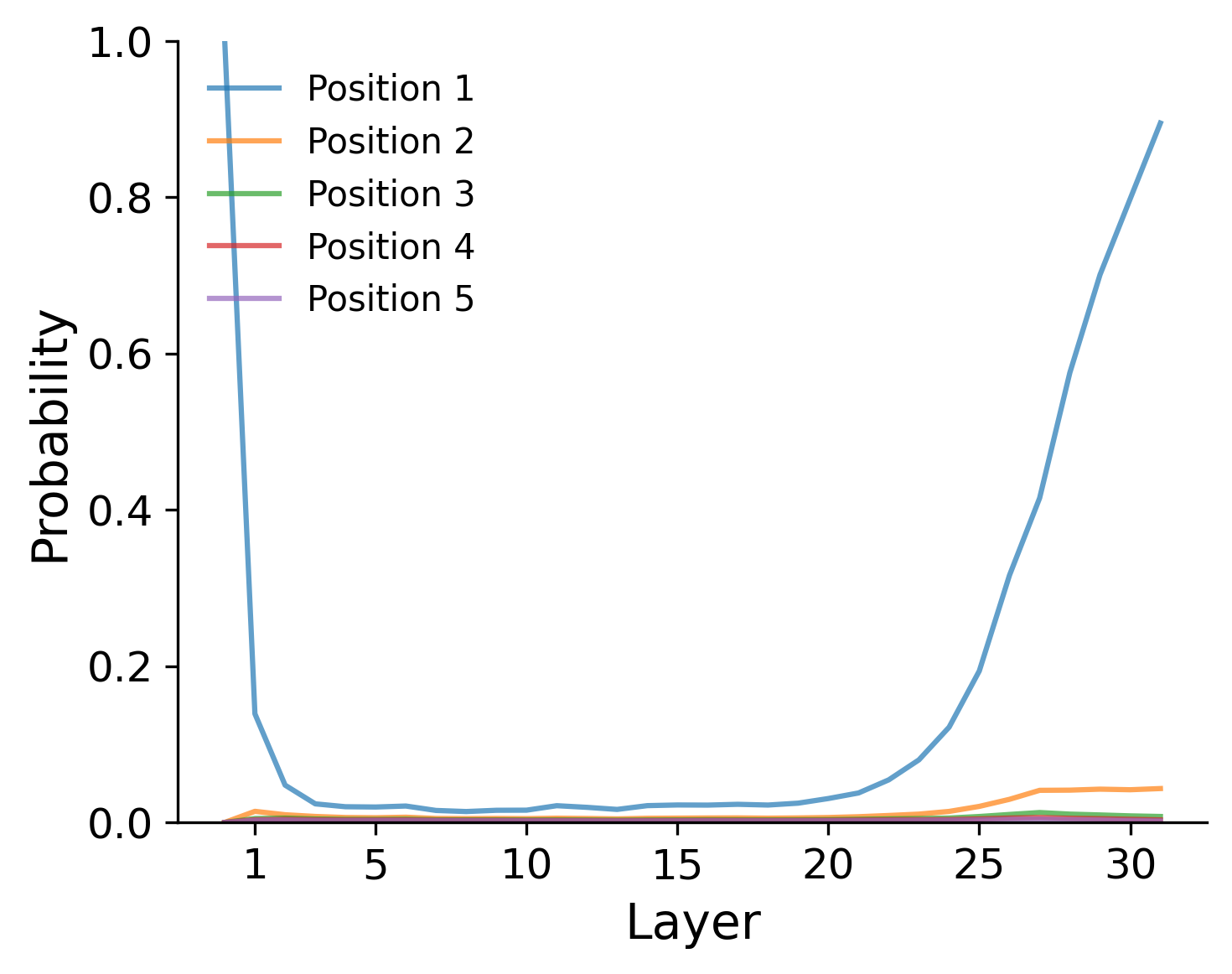}
        \caption{\whisper{}}
        \label{fig:whisper_topk}
    \end{subfigure}
    \caption{Probability of the top 5 tokens across layers for (a)~\qwen{} and (b)~\whisper{}.}
    \label{fig:logit_lens_prob_top_5}
\end{figure}

Figure~\ref{fig:logit_lens_prob_top_5} presents how the probabilities of the top-5 tokens evolve across layers. Here, position 1 corresponds to the token with the highest probability at that layer, position 2 to the next highest, and so on.

In layer 1, \whisper{} is highly confident: the most probable token receives a probability of 1. However, this confidence rapidly declines, and the probabilities across all positions quickly approach zero. Around layer 22, as shown earlier, the probability of the top-ranked token begins to rise again, eventually reaching an average of about 0.8, while the probabilities for positions~2–5 remain very low, staying close to zero.

In contrast, \qwen{} exhibits a different pattern. Throughout most layers, the top-ranked token receives the highest probability, but only around 0.2. From approximately layer 20 onward, this probability starts to increase, also stabilizing around 0.8 in the final layers. Unlike \whisper{}, however, the token in position 2 in \qwen{} maintains a probability of about 0.1 from around layer~20 onward.


\begin{figure}[h]
    \centering
    \begin{subfigure}{0.8\linewidth}
        \centering
        \includegraphics[width=\linewidth]{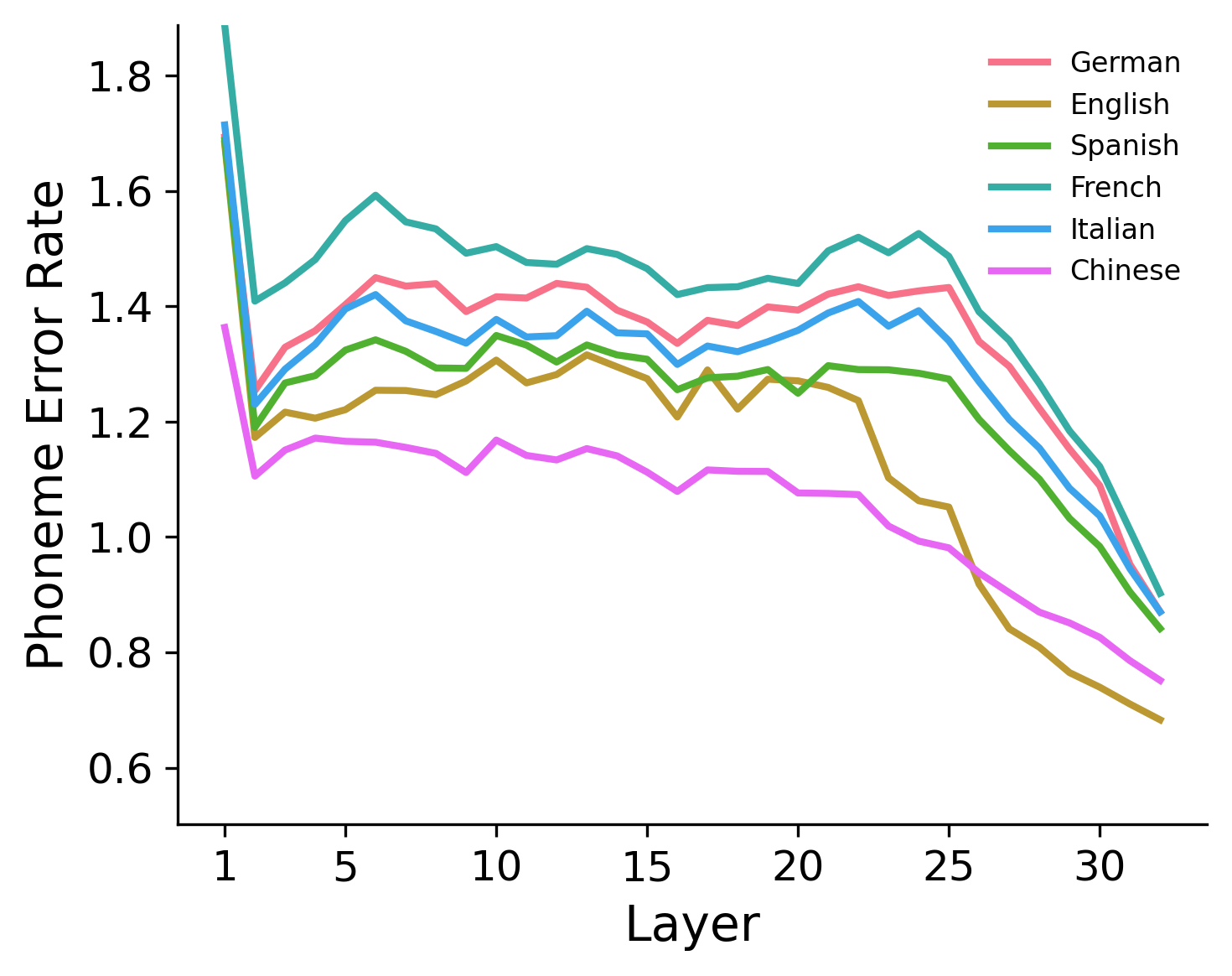}
        \caption{\qwen{}}
        \label{fig:qwen_per}
    \end{subfigure}
    \vspace{0.5em}
    \begin{subfigure}{0.8\linewidth}
        \centering
        \includegraphics[width=\linewidth]{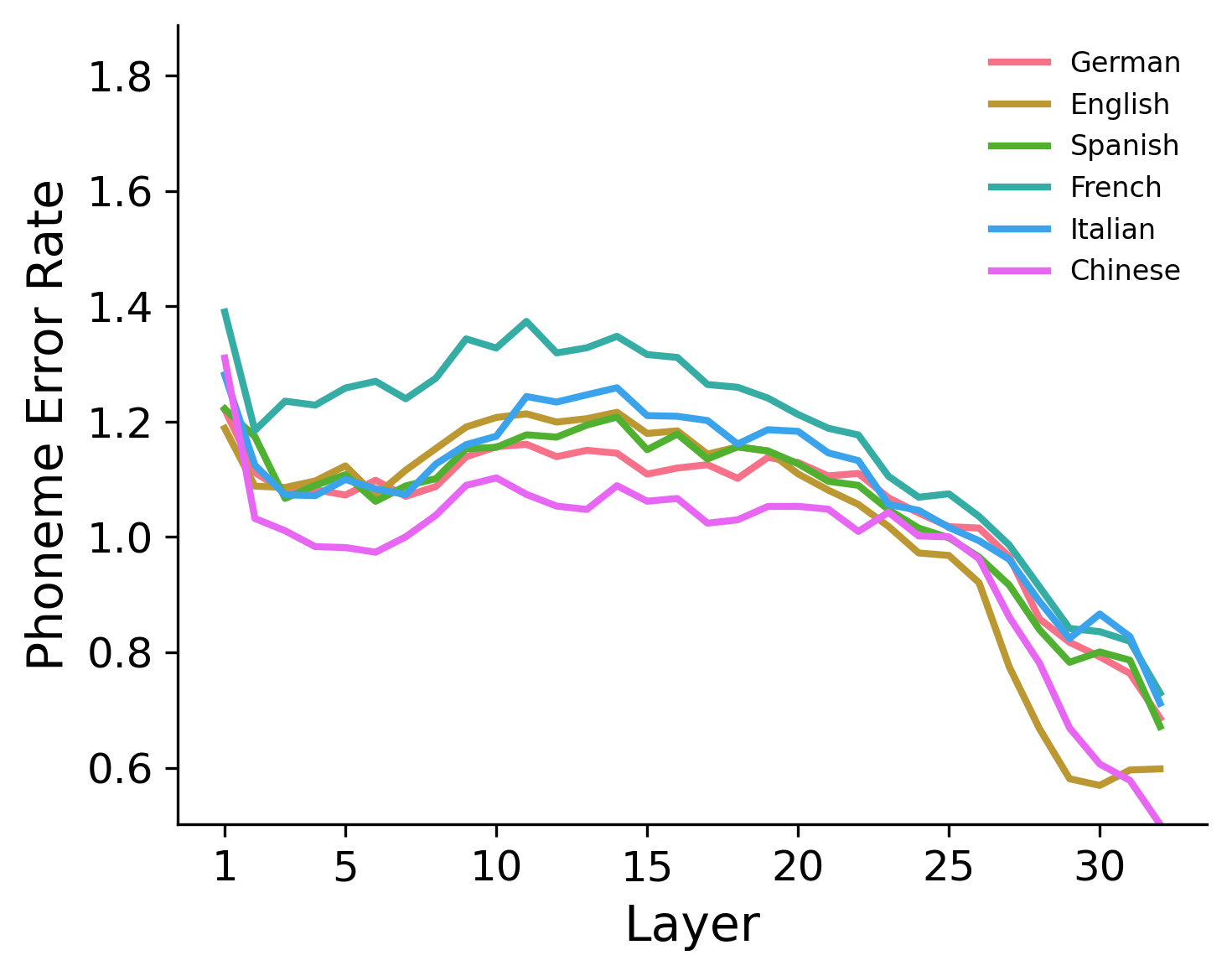}
        \caption{\whisper{}}
        \label{fig:whisper_per}
    \end{subfigure}
    \caption{Phoneme Error Rate by layer and language for (a)~\qwen{} and (b)~\whisper{}.}
    \label{fig:per_language}
\end{figure}

\textbf{Phoneme Error Rate (PER).} We propose PER as an alternative to Word Error Rate (WER), computed at the phoneme level. Instead of applying a uniform penalty, substitutions between phonemes from the same family (e.g., vowels or nasals) receive a reduced penalty of 0.5 to better reflect phonetic closeness. Each token is first converted to its phoneme sequence, and tokens without acoustic content (such as punctuation or language tags) are excluded from the calculation for both \whisper{} and \qwen{}.
Deriving phonemes first requires detecting the language of each token, which is challenging given the short length of tokens. We apply a set of simple rules: tokens containing characters from the CJK Unified Ideographs range are labeled as Chinese (even though this range also includes Japanese kanji and Korean hanja, we apply this rule because our data only include Chinese audio); tokens containing Cyrillic characters are labeled as Russian; and all other tokens are classified using fastText’s language detection \cite{bojanowski2017enriching}. Finally, we use Phonemizer \cite{Bernard2021} to generate the corresponding phoneme sequences for each token based on its detected language.
Figure~\ref{fig:per_language} shows the acoustic distance, measured by phoneme error rate (PER), between the final selected token and the top five candidate tokens produced by the models at each layer, broken down by language. For both models, PER starts very high at the first layer, then remains relatively stable until around layer~21, after which it drops sharply. Overall, \whisper{} achieves lower PER across all languages, and both models perform best (i.e., show lower PER) on English and Chinese.

\begin{figure}[h]
    \centering
    \begin{subfigure}{0.9\linewidth}
        \centering
        \includegraphics[width=\linewidth]{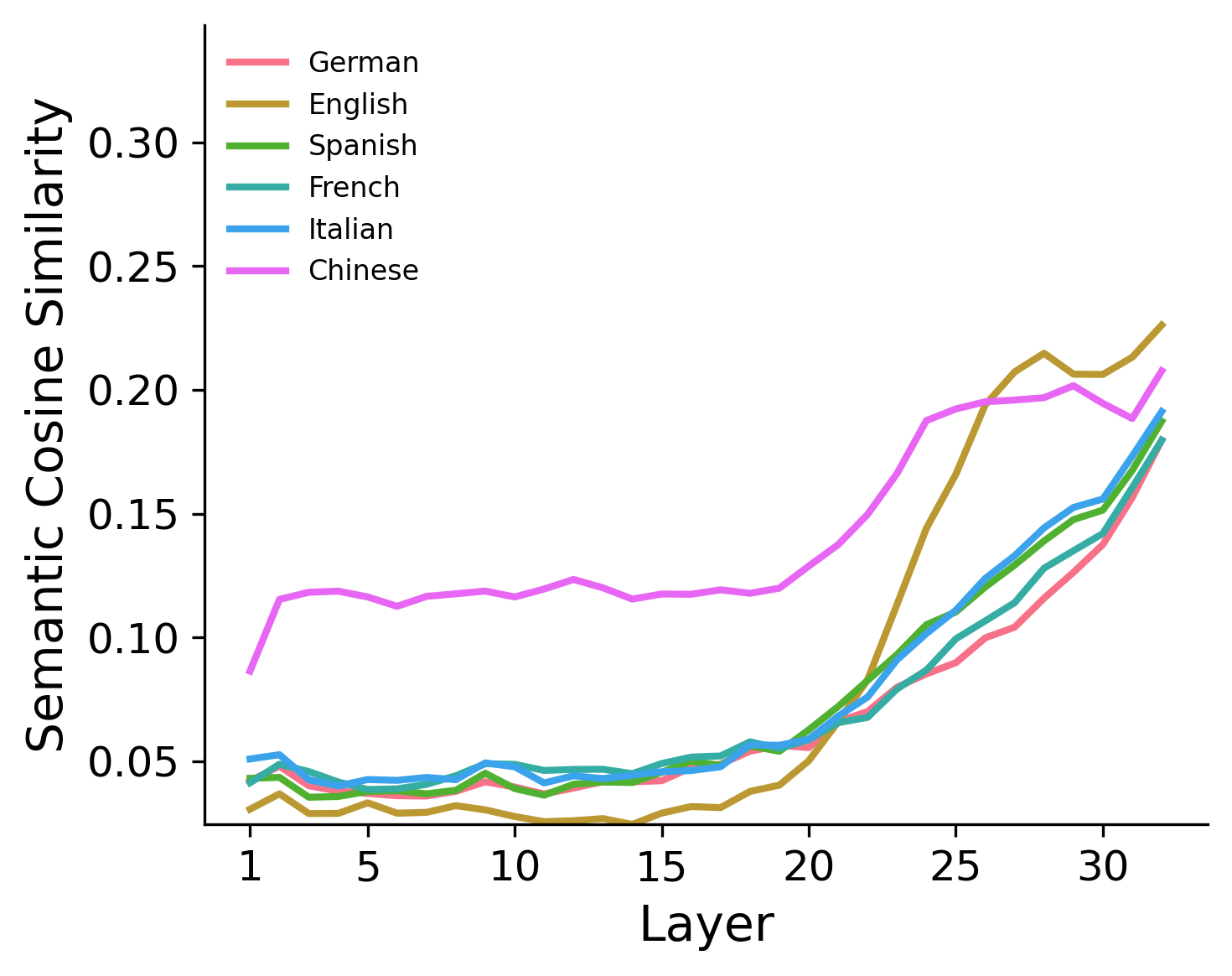}
        \caption{\qwen{}}
        \label{fig:qwen_semantic}
    \end{subfigure}
    \vspace{0.5em}
    \begin{subfigure}{0.9\linewidth}
        \centering
        \includegraphics[width=\linewidth]{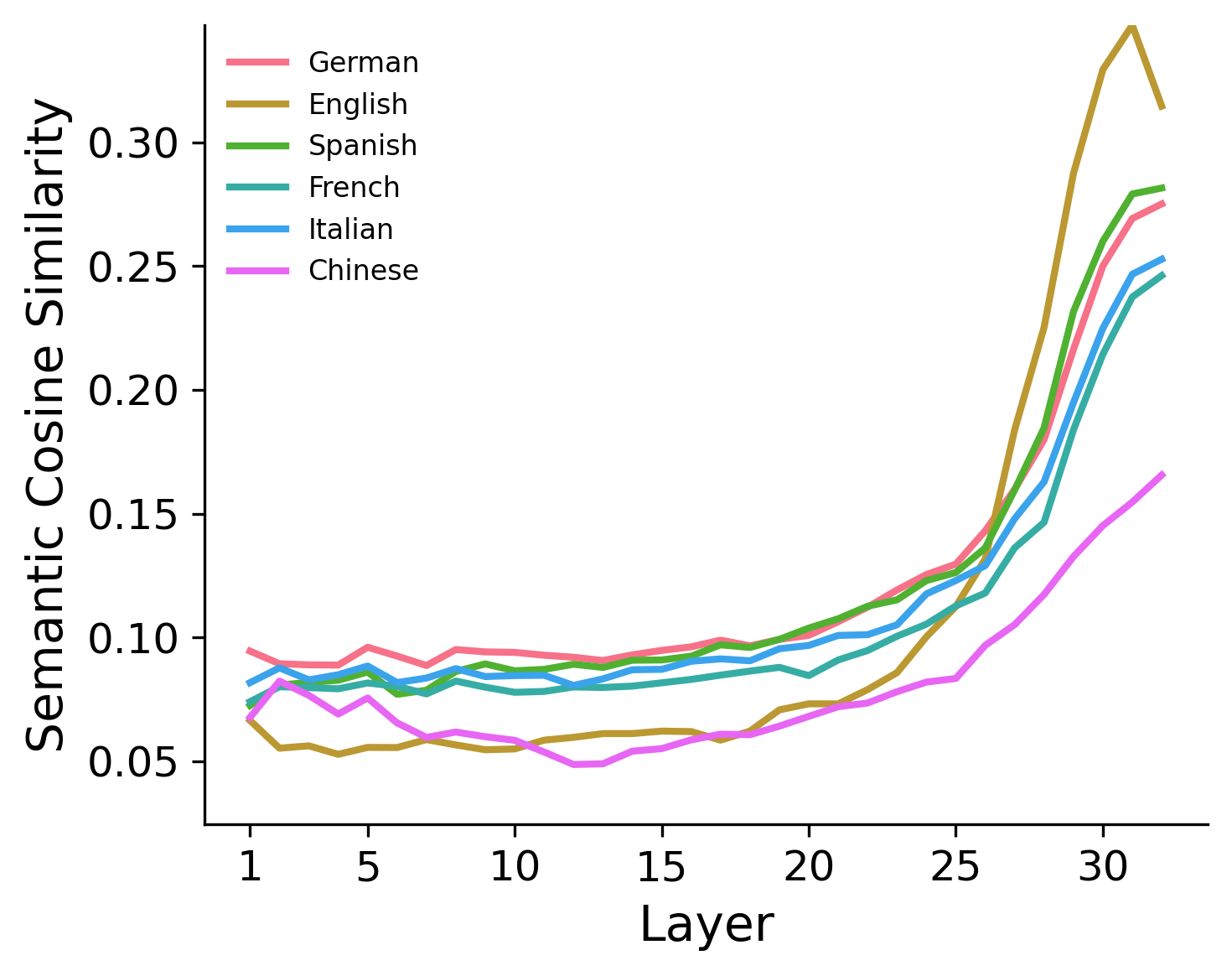}
        \caption{\whisper{}}
        \label{fig:whisper_semantic}
    \end{subfigure}
    \caption{Tokens semantic cosine similarity by layer and language for (a)~\qwen{} and (b)~\whisper{}.}
    \label{fig:semantic_lang}
\end{figure}


\textbf{Semantic similarity}. We compute semantic embeddings for each token using fastText \cite{bojanowski2017enriching} and measure their cosine similarity to the final selected token. The embedding model is selected according to the language of the audio. We exclude language tokens and other special tokens (i.e., tokens in the <|value|> format), but retain punctuation tokens since they can carry semantic meaning.
Figure~\ref{fig:semantic_lang} shows the average cosine similarity between the final selected token and the top five candidates across layers and languages. The \whisper{} model displays an interesting pattern: English starts with relatively low similarity in earlier layers but reaches the highest similarity in later layers, while Chinese consistently shows the lowest similarity. In contrast, \qwen{} shows higher similarity for Chinese already in earlier layers and again in the final layers, where English and Chinese lead. Notably, semantic similarity scores in \qwen{} are overall lower than in \whisper{}. The stronger performance on Chinese in \qwen{} may reflect its training on large amounts of Chinese data, whereas \whisper{}'s weaker results for Chinese could indicate less training data in that language.

\begin{figure*}
    \centering
    \includegraphics[width=0.9\linewidth]{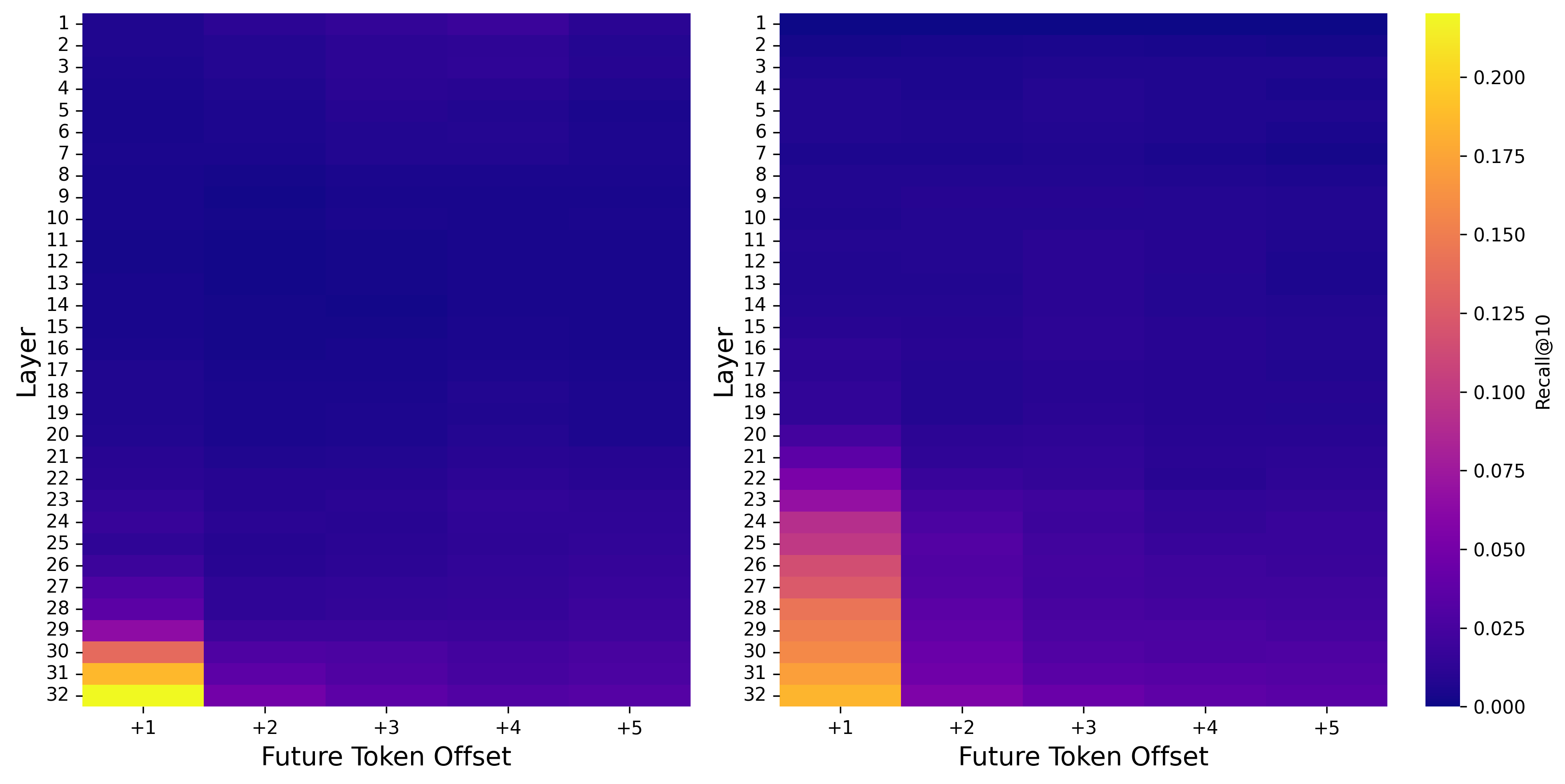}
    \caption{Future token prediction heatmap showing Recall@10 across layers and offsets $i$ for \whisper{} (left) and \qwen{} (right)} 
    \label{fig:future}
\end{figure*}

\begin{figure*}
    \centering
    \includegraphics[width=0.9\linewidth]{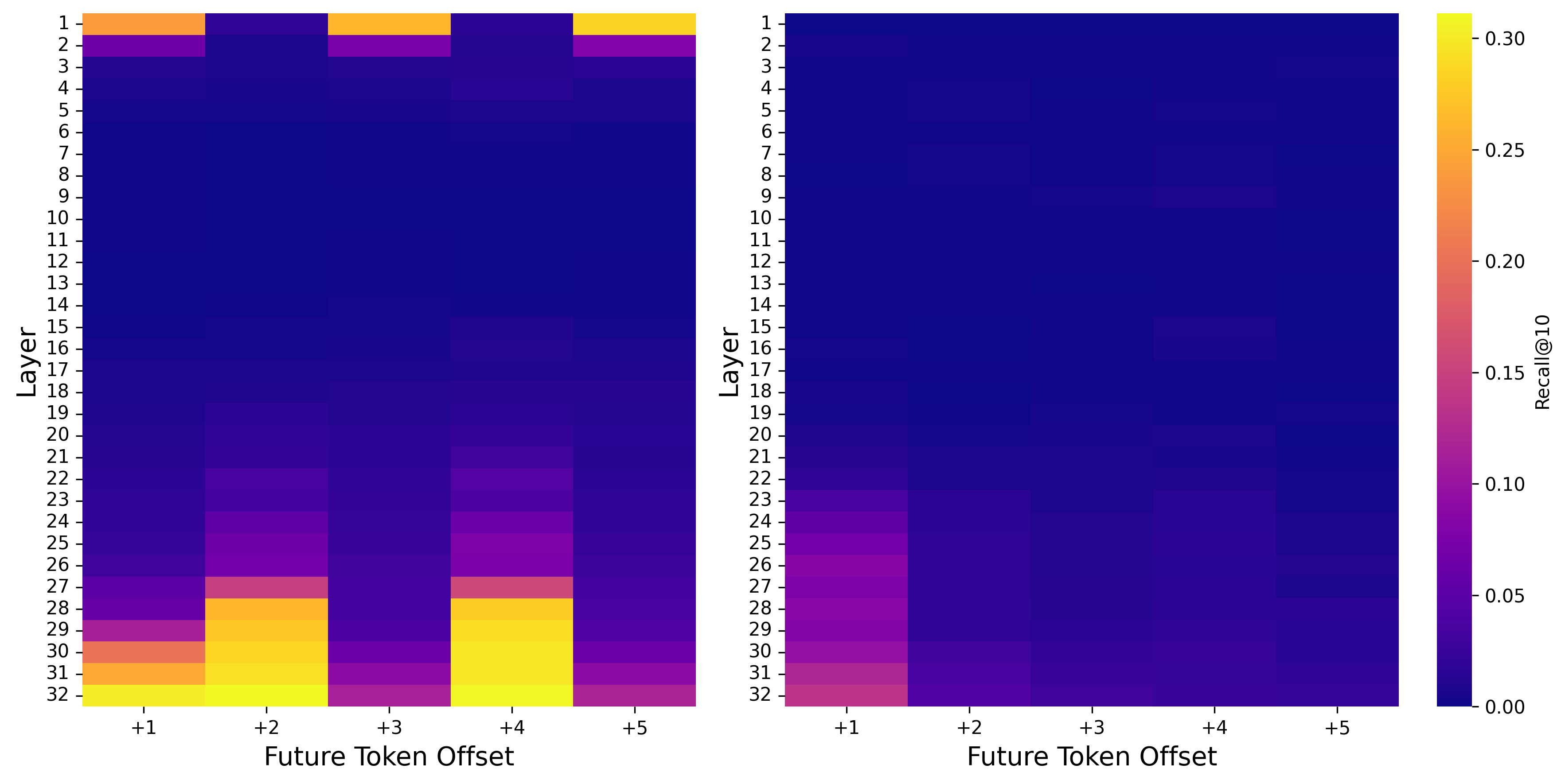}
    \caption{Japanese future token prediction heatmap showing Recall@10 across layers and offsets $i$ for \whisper{} (left) and \qwen{} (right)} 
    \label{fig:future_ja}
\end{figure*}

\paragraph{Next-Token Prediction Capabilities}

To better understand the capacity of the ASR model, we evaluate its ability to predict multiple positions beyond the next immediate token at position $s$. We consider tokens at positions $s+i$ for all  $1 \leq i \leq 5$. For each offset  $i$, we measure how often the ground-truth token at position $s+i$  appears among the model's top 10 predicted tokens at the corresponding layer. We use a metric we call Recall@10 to quantify the fraction of instances in which the correct future token is included within the top-ranked predictions. 
Figure~\ref{fig:future} presents the results for both \whisper{} and \qwen{}. As shown, \qwen{} achieves a relatively high Recall @ 10 for $s+1$ (the immediate next token) starting from approximately layer~21, indicating that these deeper layers already encode useful information about the upcoming token. Furthermore, even for $s+2$, \qwen{} maintains some predictive capability from the same layer onward, suggesting that it can anticipate tokens slightly further in the sequence.
In contrast, \whisper{} exhibits a markedly different pattern. For $s+1$, the model shows a notable rise in Recall@10 only starting around layer~29, although this increase is steeper than that observed in the \qwen{} model. Similar to \qwen{}, \whisper{} also demonstrates some prediction capability for $s+2$.

Figure~\ref{fig:future_ja} presents the next-token prediction capabilities of \whisper{} and \qwen{} on Japanese. Unlike other languages, \whisper{} shows an alternating pattern for Japanese: at even offsets~$i$, the model achieves higher Recall@10 scores starting from around layer 27; whereas at odd offsets$i$, its prediction capability is generally weaker in deeper layers but unexpectedly higher at layer~1. This is a particularly interesting behavior, as it does not appear in the \qwen{} model.

\section{Whisper's Encoder Understands Semantic}
\label{appendix:semantic}

This appendix provides detailed results from our linear probe experiments across 47 semantic category pairs on Whisper Large V3 encoder layers 22-31.

\subsection{Experimental Setup}

Our synthetic audio dataset comprises terms from 11 semantic categories: Animals, Tools, Fruits, Professions, Clothing, Countries, Musical Instruments, Body Parts, Weather, Transportation, and Academic Subjects. Each category contains 50 terms synthesized with consistent voice parameters. We train binary linear classifiers on 768-dimensional encoder representations using L2 regularization ($\lambda = 0.01$) and evaluate on held-out test sets.


\begin{table}[h!]
\centering
\small
\begin{tabular}{lc}
\toprule
\textbf{Category Pair} & \textbf{Accuracy} \\
\midrule
Countries vs. Tools & 100.0\% \\
Countries vs. Clothing & 100.0\% \\
Musical Instr. vs. Countries & 96.7\% \\
Countries vs. Weather & 93.3\% \\
\bottomrule
\end{tabular}
\caption{Top Performing Category Pairs (Layer 31)}
\label{tab:top_pairs}
\end{table}

\begin{table}[h!]
\centering
\small
\begin{tabular}{lc}
\toprule
\textbf{Category} & \textbf{Average Accuracy} \\
\midrule
Countries & 91.2\% \\
Professions & 84.2\% \\
Body Parts & 82.7\% \\
Academic Subjects & 73.0\% \\
Clothing & 72.6\% \\
\bottomrule
\end{tabular}
\caption{Average Category Performance (Layer 31)}
\label{tab:category_avgs}
\end{table}





\subsection{Visual Analysis}

Figure~\ref{fig:semantic_progression} illustrates semantic classification progression for selected high-performing category pairs. 
Tables~\ref{tab:top_pairs} and~\ref{tab:category_avgs} summarize these results, showing perfect separation for several category pairs and consistently higher averages for \textit{Countries} compared to other categories.

\begin{figure*}[h!]
\centering
\includegraphics[width=1.0\linewidth]{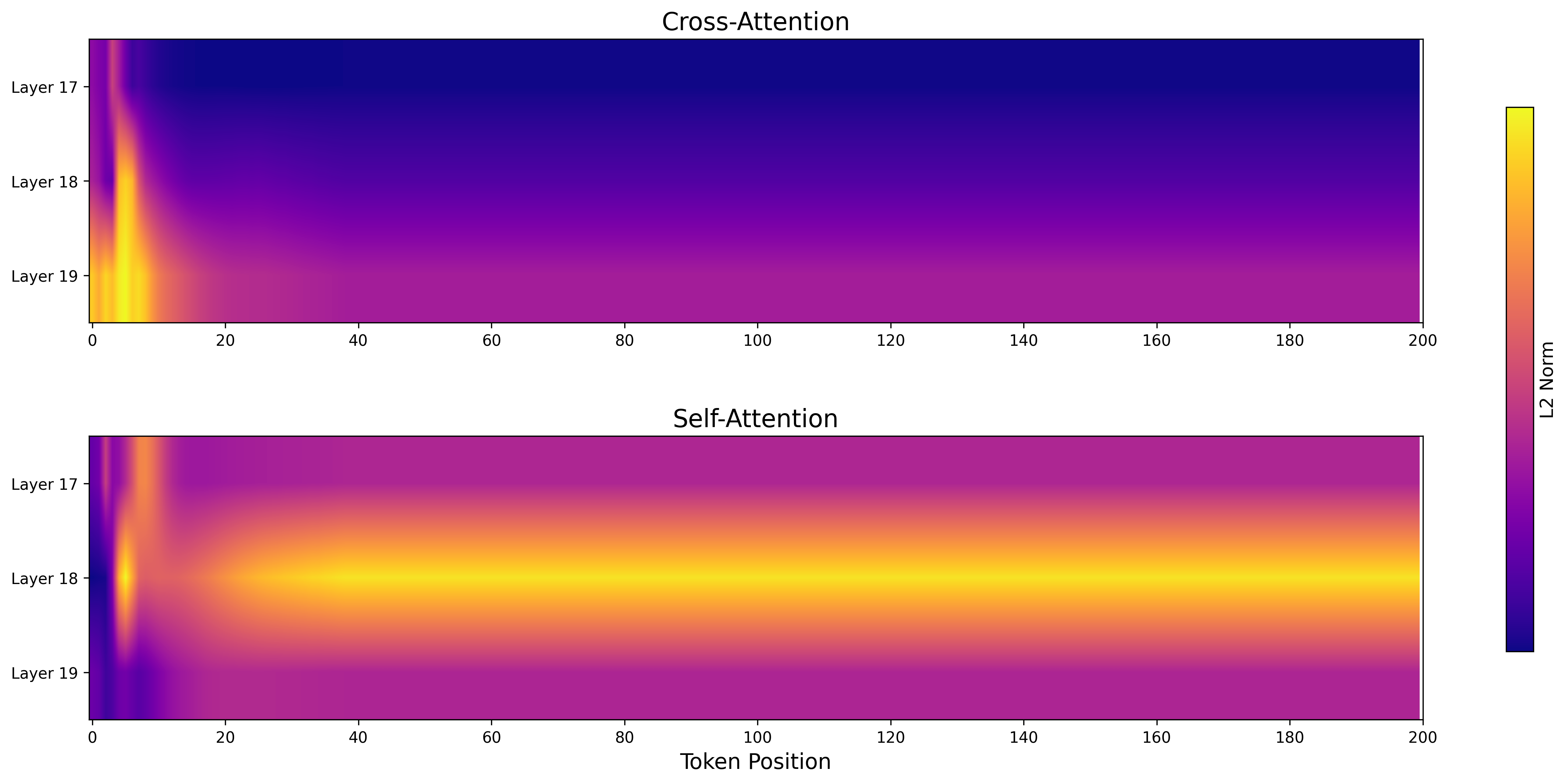}
\caption{Attention component L2 norm traces during repetitive sequence generation. Cross-attention (top) shows peak-then-decline pattern while self-attention (bottom) exhibits sustained elevation, with Layer 18 demonstrating the most pronounced dual signature.}
\label{fig:layer18_disconnection}
\end{figure*}

\section{Repetitions}
\subsection{Layer 18 Norm Patterns During Repetition}

Our analysis reveals distinctive activation patterns in Layer 18 during repetitive sequences that provide insights into the underlying mechanisms of repetition generation. Figure~\ref{fig:layer18_disconnection} shows L2 norm traces for cross-attention and self-attention components across layers 17, 18, and 19 during a repetitive sequence.

\paragraph{Cross-Attention as Repetition Trigger}
Layer 18's cross-attention component shows intense initial activation followed by sustained decline (Figure~\ref{fig:layer18_disconnection}, top panel). This pattern suggests that cross-attention initially attempts to maintain alignment with the audio source but progressively loses this connection as repetition begins. 

The declining cross-attention norms indicate a triggering mechanism where the encoder-decoder alignment breaks down, initiating the repetitive state.

\paragraph{Self-Attention as Repetition Manifestation}  
The simultaneous elevation in self-attention norms that persists throughout repetition (Figure~\ref{fig:layer18_disconnection}, bottom panel) reflects the decoder's internal repetitive processing. 

This sustained activity pattern manifests the pathological state where the decoder becomes internally focused on repetitive patterns rather than attending to new audio input. The persistent elevation suggests that self-attention captures and maintains the repetitive behavior once initiated.

\paragraph{Intervention Implications}
The effectiveness of cross-attention patching versus self-attention intervention failure can be understood through this dual-component view. 

We hypothesize that cross-attention interventions target the trigger mechanism, preventing the initial breakdown in audio alignment. In contrast, self-attention interventions attempt to modify the manifestation after the repetitive state has already been established, proving less effective. Layer 18's distinctive signature in both components (Figure~\ref{fig:layer18_disconnection}) explains why interventions at this specific layer are most effective for repetition control. 

Further investigation into this trigger-manifestation mechanism could provide deeper insights into the architectural foundations of hallucination control in ASR models.


\begin{figure*}[h!]
\centering
\includegraphics[width=1.05\linewidth]{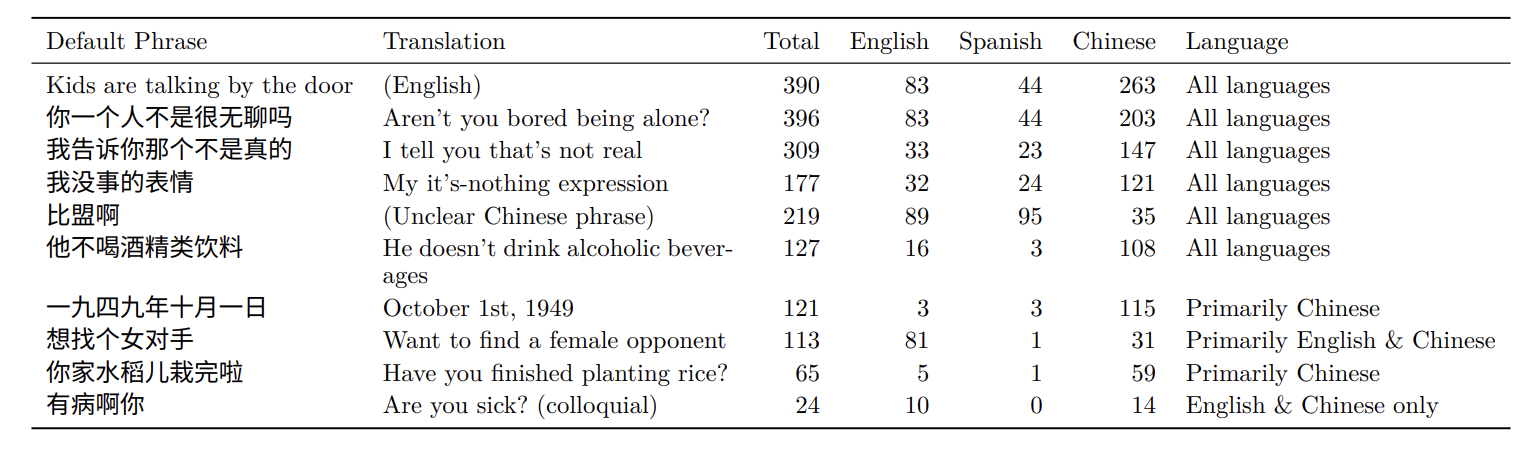}
\caption{Examples of Frequent Default Phrases in \qwen{}}
\label{tab:defaults}
\end{figure*}


\newpage
\subsection{Types Of Repetitions}

\begin{itemize}
    \item \textbf{Code Switching:} 
    \\
    "The genus name comes from the classical Latin word co-co-co-co-co-co-co-co-co-co-co-co-co-co-co-co-co-co-co-co-co-co-co-co-co-co-co-co-co-co-co-co-co-co-co-co-co-co-co-co-co-co-co-co-co-co-co-co-co-co-co-co-co-co-co-co-co-co-co-co-co-co-co-co-co-co-co-co-co-co-co-co-co-co-co-co-co-co-co-co-co-co-co-co-co-co-co-co-co-co-co-co-co-co-co-co-co-co-co-co-co-co-co-co-co-co-co-co-co-co-co-co-co-co-co-co-co-co-co-co-co-co-co-co-co-co-co-co-co-co-co-co-co-co-co-co-co-co-co-co-co-co-co-co-co-co-co-co-co-co-co-"

    \item \textbf{Repetitive Audio:} In the acoustic the word "ha" was said 5 times: 
    
    "Ha ha ha ha ha ha ha ha ha ha ha ha ha ha ha ha ha ha ha ha ha ha ha ha ha ha ha ha ha ha ha ha ha ha ha ha ha ha ha ha ha ha ha ha ha ha ha ha ha ha ha ha ha ha ha ha ha ha ha ha ha ha ha ha ha ha ha ha ha ha ha ha ha ha ha ha ha ha ha ha ha ha ha ha ha ha ha ha ha ha ha ha ha ha ha ha ha ha ha ha ha ha ha ha ha ha ha ha ha ha ha ha ha ha ha ha ha ha ha ha ha ha ha ha ha ha ha ha ha ha ha ha ha ha ha ha ha ha ha ha ha ha ha ha ha ha ha ha ha ha ha ha ha ha ha ha ha ha ha ha ha ha ha ha ha ha ha ha ha ha ha ha ha ha ha ha ha ha ha ha ha ha ha ha ha ha ha ha ha ha ha ha ha ha ha ha ha ha ha ha ha ha ha ha ha ha ha ha ha ha ha ha ha ha ha ha ha ha ha ha ha ha ha ha ha ha ha ha ha ha ha ha ha ha ha ha ha ha ha ha ha ha ha ha ha ha ha ha ha ha ha ha ha ha ha ha ha ha ha ha ha ha ha ha ha ha ha ha ha ha ha ha ha ha ha ha ha ha ha ha ha ha ha ha ha ha ha ha ha ha ha ha ha ha ha ha ha ha ha ha"
    
    \item \textbf{Short Noise:} 
    \\
    "Rrrrrrrrrrrrrrrrrrrrrrrrrrrrrrrrrrrrrrrrrrrrrrrrrr
    rrrrrrrrrrrrrrrrrrrrrrrrrrrrrrrrrrrrrrrrrrrrrrrrrrrr
    rrrrrrrrrrrrrrrrrrrrrrrrrrrrrrrrrrrrrrrrrrrrrrrrrrrr
    rrrrrrrrrrrrrrrrrrrrrrrrrrrrrrrrrrrrrrrrrrrrrrrrrrrr
    rrrrrrrrrrrrrrrrrrrrrrrrrrrrrrrrrrrrrrrrrrrrrrrrrrrr
    rrrrrrrrrrrrrrrrrrrrrrrrrrrrrrrrrrrrrrrrrrrrrrrrrrrr
    rrrrrrrrrrrrrrrrrrrrrrrrrrrrrrrrrrrrrrrrrrrrrrrrrrrr
    rrrrrrrrrrrrrrrrrrrrrrrrrrrrrrrrrrrrrrrrrrrrrrrrrrrr
    rrrrrrrrrrrrrrrrrrrrrrrrrrrrrrrrrrrrrrrrrrrrrrrrrrrr
    rrrrrrrrrrrrrrrrrrrrrrrrrrrrrrrrrrrrrrrrrrrrrrrrrrrr
    rrrrrrrrrrrrrrrrrrrrrrrrrrrrrrrrrrrrrrrrrrrrrrrrrrrr
    rrrrrrrrrrrrrrrrrrrrrrrrrrrrr"
\end{itemize}

\section{Encoder Lens}

\begin{figure*}[h!]
\centering
\includegraphics[width=1.0\linewidth]{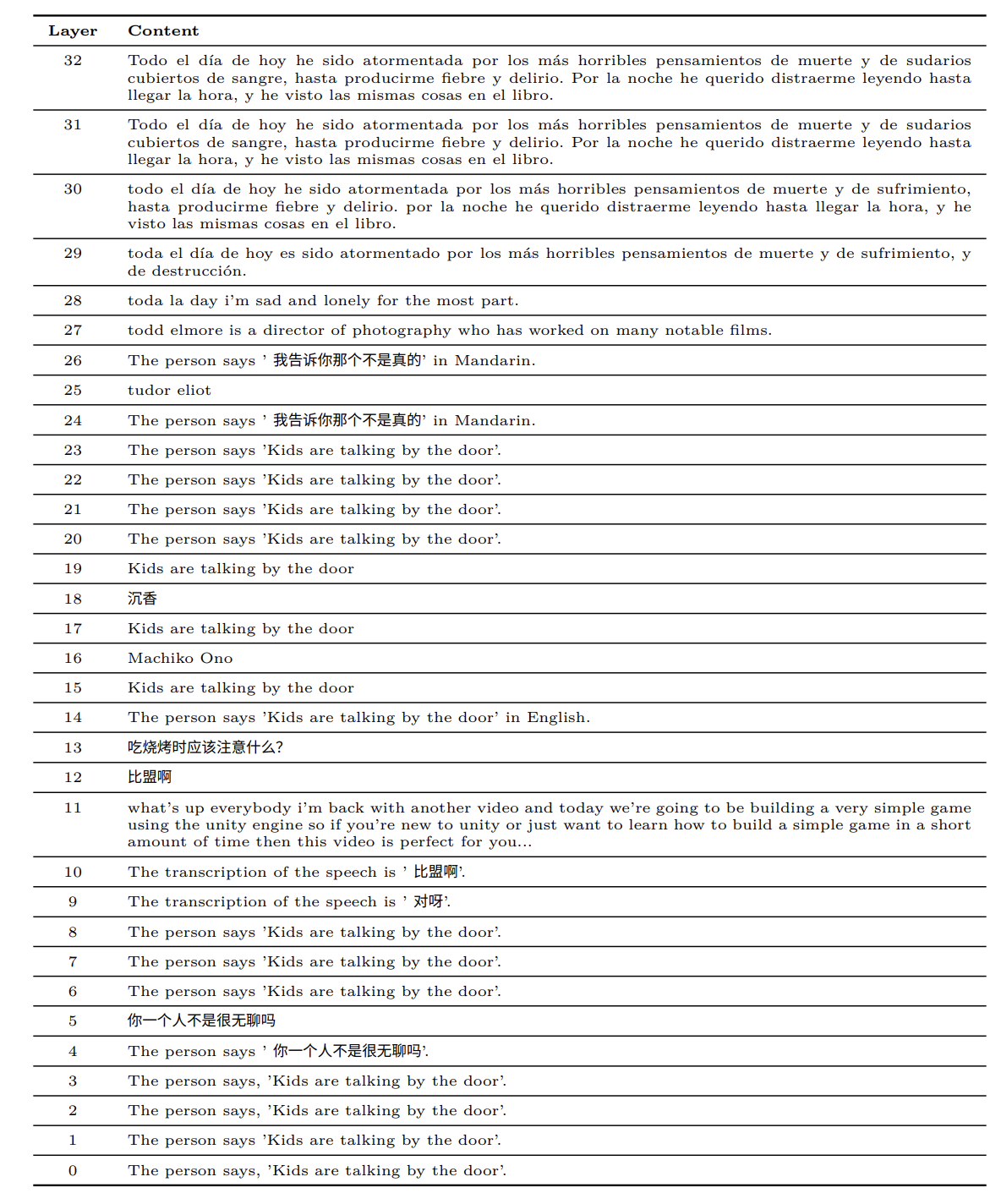}
\caption{Spanish Example - \qwen{}}
\label{fig:spanish}
\end{figure*}

\begin{figure*}[h!]
\centering
\includegraphics[width=1.0\linewidth]{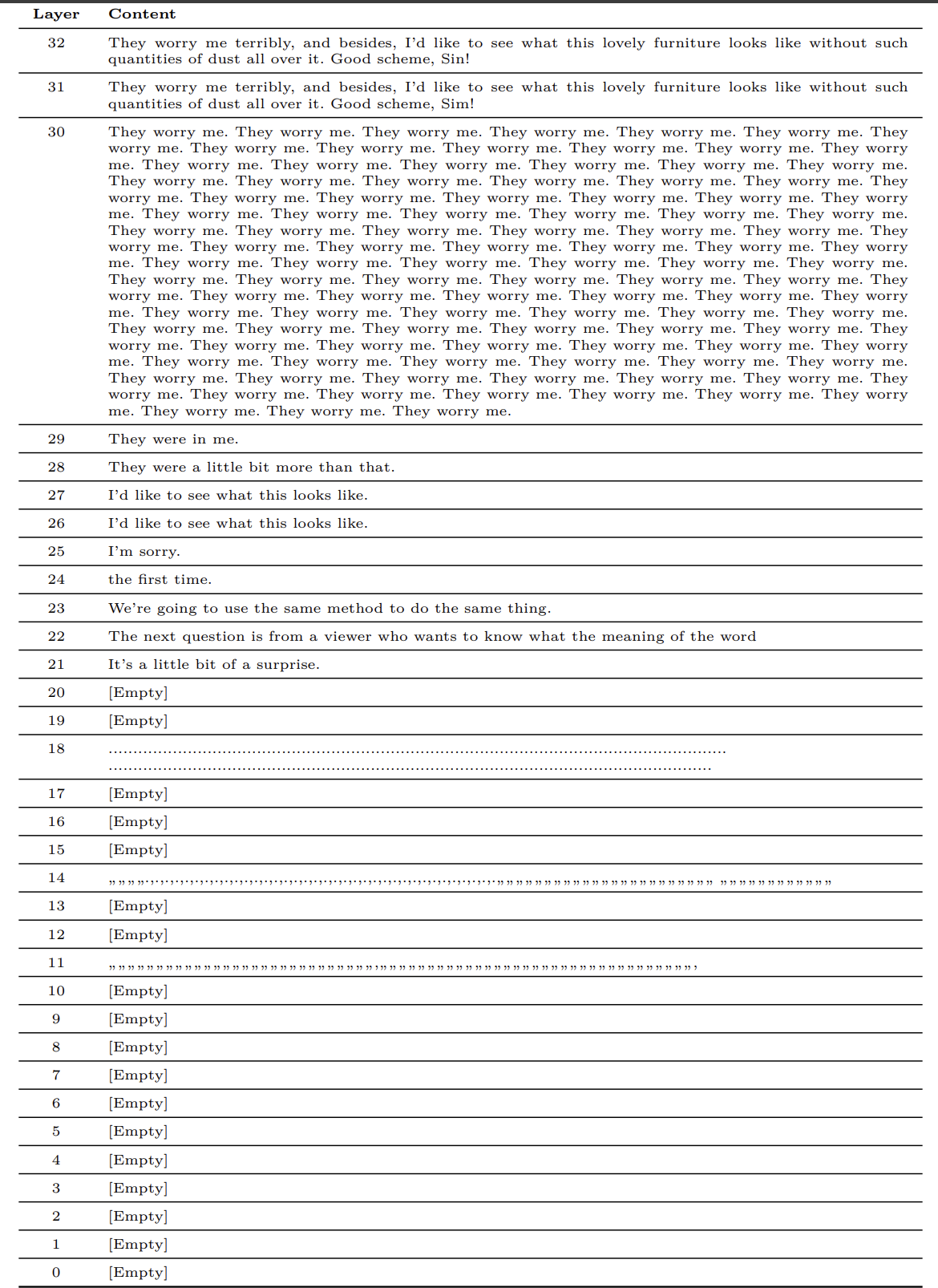}
\caption{English Example - Whisper}
\label{fig:eng}
\end{figure*}

\begin{figure*}[h!]
\centering
\includegraphics[width=1.0\linewidth]{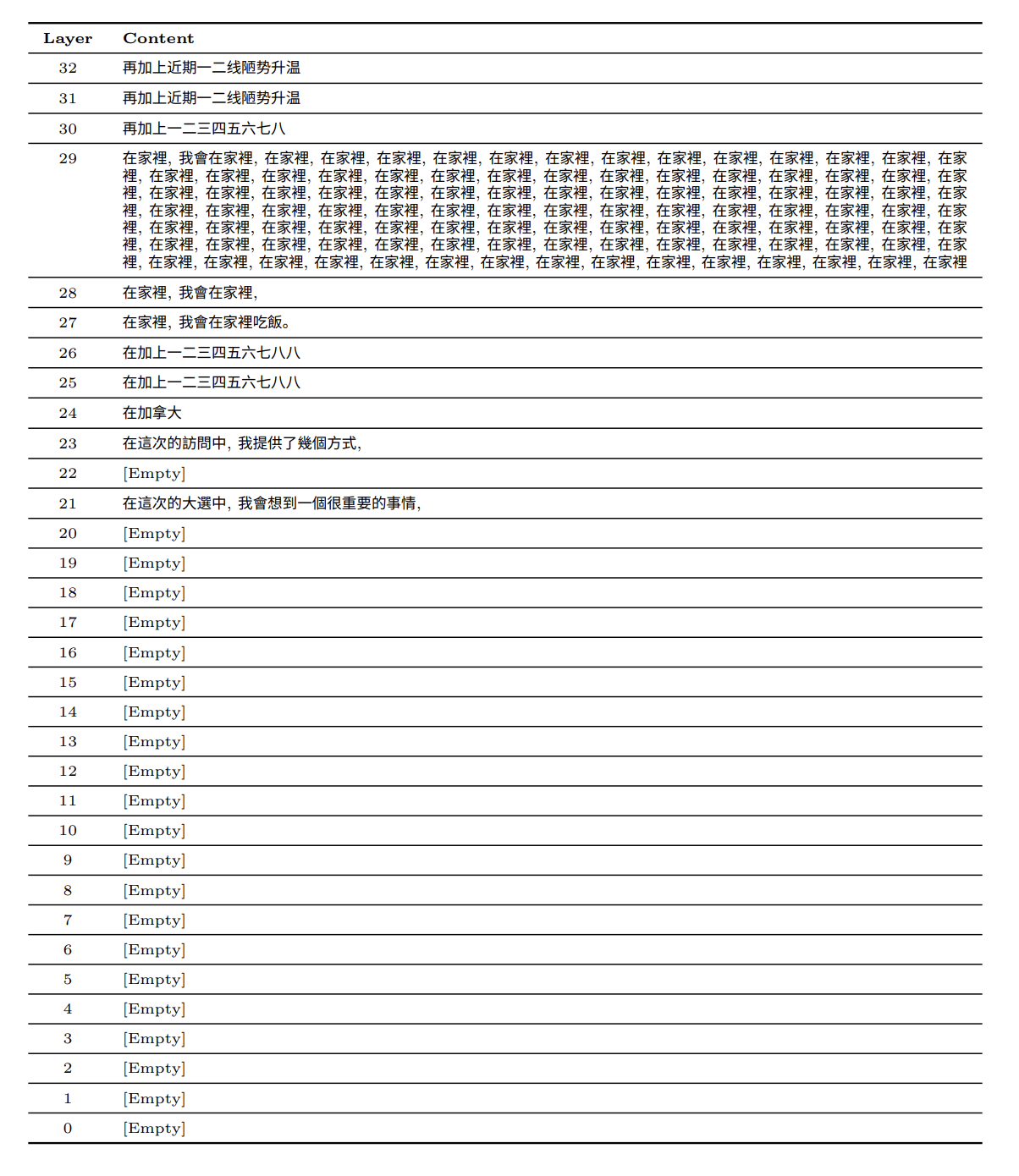}
\caption{Chinese Example - Whisper}
\label{fig:zh}
\end{figure*}

We provide qualitative examples of decoder outputs from different encoder layers using the Whisper or \qwen{} models, across three languages: Spanish, English, and Chinese, shown in Figures \ref{fig:spanish}, \ref{fig:eng} and \ref{fig:zh} respectively.

\clearpage

\end{document}